\pdfoutput=1
\documentclass[aps,prl,reprint,showpacs,superscriptaddress]{revtex4-2}

\usepackage{comment}

\usepackage[colorlinks,urlcolor=blue,citecolor=blue]{hyperref}
\usepackage[rgb,dvipsnames]{xcolor}
\usepackage[export]{adjustbox}
\usepackage{graphicx}
\usepackage{bbold}
\usepackage{qcircuit}
\usepackage{braket}
\usepackage{amsmath,amssymb}
\usepackage[per-mode=symbol]{siunitx}
\usepackage{changepage}
\usepackage{algorithm}        
\usepackage{algorithmic} 
\DeclareSIUnit\angstrom{\text{\AA}}

\newcommand{\numfreqs}{N_{\omega}}

\usepackage{mathtools}

\newcommand{\papertitle}{Estimating Green's functions with a robust quantum Arnoldi method}

\begin{document}

\title{\papertitle}
\author{Jacob S. Nelson}
\email{jnelso2@sandia.gov}
\affiliation{Center for Quantum Information and Control (CQuIC),
             University of New Mexico,
             Albuquerque, NM}
\affiliation{Quantum Algorithms and Applications Collaboratory (QuAAC), Sandia National Laboratories, Albuquerque, NM}

\author{Andrew D. Baczewski}
\email{adbacze@sandia.gov}
\affiliation{Center for Quantum Information and Control (CQuIC),
             University of New Mexico,
             Albuquerque, NM}
\affiliation{Quantum Algorithms and Applications Collaboratory (QuAAC), Sandia National Laboratories, Albuquerque, NM}

\date{\today}

\begin{abstract}
Many applications of Green's functions (GFs) require their evaluation over intervals or at multiple points, motivating quantum algorithms that return an efficiently computable functional representation rather than mere point estimates. 
We introduce a robust quantum Arnoldi method (ROQAM) that achieves this goal. 
Its robustness is derived from formulation in terms of orthogonal polynomials, which preserves the upper-Hessenberg structure of the projected matrices despite finite-precision estimation. 
We also show that as the iteration depth increases, the precision required for matrix-element estimation can be reduced. 
Resource estimates for the spectral function of a quantum impurity model indicate that ROQAM outperforms pointwise estimation via quantum singular value transformation by multiple orders of magnitude. 
Finally, we show that the ROQAM can be used to estimate GFs at nonzero temperatures using only a single Krylov subspace.
\end{abstract}

\maketitle

\textit{Introduction.--- } Green's functions (GFs) are widely used in quantum many-body physics~\cite{fetter2012quantum}, as they relate to many diverse responses measured in experiments ranging from microscopies~\cite{schiller2000theory} to spectroscopies~\cite{golze2019gw}, in contexts ranging from nuclear reactions~\cite{dickhoff2004self} to degenerate plasmas~\cite{dornheim2023electronic}.
They are also central theoretical quantities in embedding methods, like dynamical mean-field theory (DMFT)~\cite{RevModPhys.68.13}, which has been highly effective in efficiently modeling strongly correlated systems~\cite{Kotliar_2006}.
In this Letter, we consider the problem of calculating single-particle GFs on a quantum computer.
More efficient methods for doing so will enable these powerful theoretical tools to be applied to a variety of physics simulation problems.

The quantum simulation applications that have received the most analysis involve estimating either energy eigenvalues~\cite{low2025fast} or observables associated with non-equilibrium quantum dynamics~\cite{rubin2024quantum}.
GF estimation is related to both, as the poles encode excitation energies and the functions themselves are defined as amplitudes associated with time evolution~\cite{fetter2012quantum}.
Prior work on calculating single-particle GFs on quantum computers has considered a variety of techniques, including measuring time series from real-time evolutions~\cite{PhysRevX.6.031045}, quantum linear systems solvers~\cite{Tong_2021,ralli2023calculating}, a quantum Krylov algorithm~\cite{kirby2025quantumkrylovalgorithmszego}, and even a variational algorithm~\cite{rungger2019dynamical}.
Related work includes methods for calculating higher-order response functions, as well~\cite{rall2020quantum,kharazi2025efficient}.

Most applications require knowledge of the GF on an interval of real frequencies, or even on a set of complex frequencies (e.g., Matsubara frequencies).
For this reason, Lanczos methods~\cite{lanczos} are commonly employed classically because they construct a compact representation of the GF as a function, rather than a set of point estimates.
This is accomplished by iteratively constructing an orthonormal basis from a Krylov subspace generated from the Hamiltonian $H$.
In this basis, the GF takes the form of a continued fraction with frequency-dependent coefficients~\cite{R_Haydock_1972}.
Straightforward adaptation of this method into a quantum algorithm using a block-encoding~\cite{Low_2019, Chakraborty_2019} to access $H$ results in rapid growth in the subnormalization factor with the depth of iteration.
We instead access the spectrum of $H$ through the time-evolution operator $e^{-iHt}$ and utilize the Arnoldi method~\cite{arnoldi}, the non‐Hermitian generalization of the Lanczos method.

The robust quantum Arnoldi method (ROQAM) that we propose is conceptually similar to the quantum isometric Arnoldi method in Ref.~\cite{kirby2025quantumkrylovalgorithmszego}. 
However, ours is formulated in terms of orthogonal polynomials causing the projected matrix to realize upper-Hessenberg form independent of finite-precision estimation errors.
We further show that the robustness of our method to finite-precision errors allows us to relax quantum resource requirements.
We compare resource estimates for GF estimation using the ROQAM and the quantum singular value transform (QSVT)~\cite{Gily_n_2019}, and show that the ROQAM requires fewer resources for small instances of an impurity model.
Finally, we show that it is straightforward to generalize the ROQAM to estimating GFs at nonzero temperature using the thermofield double (TFD) formalism~\cite{mann1992thermo, cottrell2019build,Harsha_2019}.

\textit{Methods.--- }
At zero temperature, the single-particle GF is defined in terms of an $n$-qubit Hamiltonian $H$ on Hilbert space $\mathcal{H}_n$ and its ground-state eigenpair $\left(\ket{\psi_0},E_0\right)$ as
\begin{subequations}
\begin{gather}
     G^+_{pq}(\omega) = \bra{\psi_0}a_{p}\left[ (\omega + i\gamma) - (H - E_0) \right]^{-1}a_{q}^{\dag}\ket{\psi_0}, \\
     G^-_{pq}(\omega) = \bra{\psi_0}a_{p}^{\dag}\left[ (\omega + i\gamma) + (H - E_0) \right]^{-1}a_{q}\ket{\psi_0}, \\
     G_{pq}(\omega) = G^+_{pq}(\omega) + G^-_{pq}(\omega).
     \label{eq:G(w)}
\end{gather}
\end{subequations}
$a_p$ ($a_p^\dagger$) is an annihilation (creation) operator acting on a suitable set of modes indexed by $p$ and $q$ and $\gamma \in \mathbb{R}^+$ is an empirical broadening parameter related to the desired spectral resolution.
The spectral function $A_{pq}(\omega)$ encodes information about single-particle excitations and it is expressed in terms of $G_{pq}(\omega)$,
\begin{equation}
    A_{pq}(\omega) = -\pi^{-1}\textrm{Im}\left[G_{pq}(\omega)\right].
\end{equation}

\begin{figure*}[!t]
    \centering
    \includegraphics[scale=0.7]{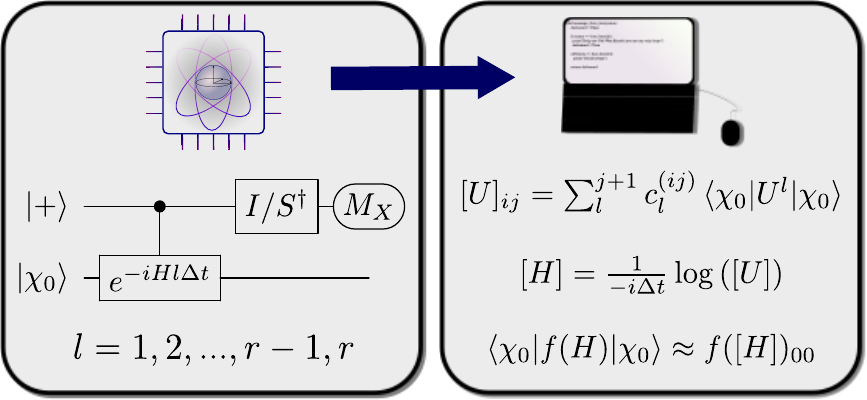}
    \caption{In the quantum Arnoldi method (ROQAM) a quantum computer is used to estimate expectation values $\braket{\chi_0|U^l|\chi_0}$ for $l = 1, 2, ..., r$. 
    These estimates are passed to a classical processor that builds $[U]$. 
    This matrix is processed into $[H]$, a representation of the Hamiltonian in the Arnoldi basis, which can be used to estimate expectation values of functions of $[H]$.}
    \label{fig:protocol}
\end{figure*}

The Arnoldi matrix function method is a two-step process to estimate the expectation value $\braket{\chi_0|f(M)|\chi_0}$. 
First, the matrix $M$ is projected into the Arnoldi basis built from the Krylov subspace $\mathcal{K}_r(M, \ket{\chi_0})$.
This is done through the Arnoldi iteration: starting with initial vector $\ket{\chi_0}$, one alternates between forming a new vector $\ket{\tilde{\chi}^{'}_j} = M\ket{\chi_{j-1}}$, then orthogonalizing $\ket{\tilde{\chi}^{'}_j}$ against all previous vectors and normalizing to form $\ket{\chi_j}$.
The projected matrix $[M]$ is then defined by the action of $M$ within the Arnoldi basis $[M]_{ij} = \braket{\chi_i|M|\chi_j}$.
This process is equivalently described as finding orthogonal polynomials that output the $(j+1)$th basis state when applied to $\ket{\chi_0}$, $P_{j+1}(M)\ket{\chi_0} = \ket{\chi_{j+1}}$.
These polynomials are defined recursively as
\begin{subequations}
\begin{gather}
    P_0(M) = 1, \\
    \Tilde{P}_{j+1}(M) = (M\!-\![M]_{jj})P_{j}(M) - \sum_{i=0}^{j-1}[M]_{ij}P_i(M), \\
    P_{j+1}(M) = [M]^{-1}_{j+1,j} \Tilde{P}_{j+1}(M).
\end{gather}
\label{eq:polys}
\end{subequations}
Likewise, the elements of $[M]$ are defined as
\begin{equation}
    [M]_{ij} = \begin{cases}
                \bra{\chi_0}P^{\dag}_i(M)MP_j(M)\ket{\chi_0}, & \text{if } i \leq j \\[0.5em]
                \sqrt{ \braket{\chi_0|\Tilde{P}^{\dag}_j(M)\Tilde{P}_j(M)|\chi_0 } }, & \text{if } i = j + 1 \\[0.5em] 
                0, & \text{if } i > j + 1. \\
             \end{cases}
    \label{eq:poly_matrix_elements}
\end{equation}
This is carried out until $j+1=r$.
Once the matrix $[M]$ is obtained, we estimate the expectation value of any matrix function as
\begin{equation}
    \braket{\chi_0|f(M)|\chi_0} \approx \braket{\chi_0|\chi_0}f([M])_{00}.
    \label{eq:A_quad}
\end{equation}
When the iteration is performed using exact arithmetic to depth $r$, the estimator error is generally upper bounded by optimal degree-$2r$ polynomial approximations to $f(M)$~\cite{ gautschi2004orthogonal, golub2009matrices,kirby2025quantumkrylovalgorithmszego}.
In practice these bounds are typically outperformed, even for finite-precision arithmetic~\cite{chen2024lanczosalgorithmmatrixfunctions}.
The projection step is much more expensive than evaluating $f([M])$, and once $[M]$ is obtained it can be reused to estimate functions of the form $f(z\mathbb{1} + M)$ for as many values of $z$ as desired~\footnote{Additionally, the same $[M]$ could be to estimate multiple different functions.}.

\begin{figure*}[!t]
    \centering
    \includegraphics[scale=0.9]{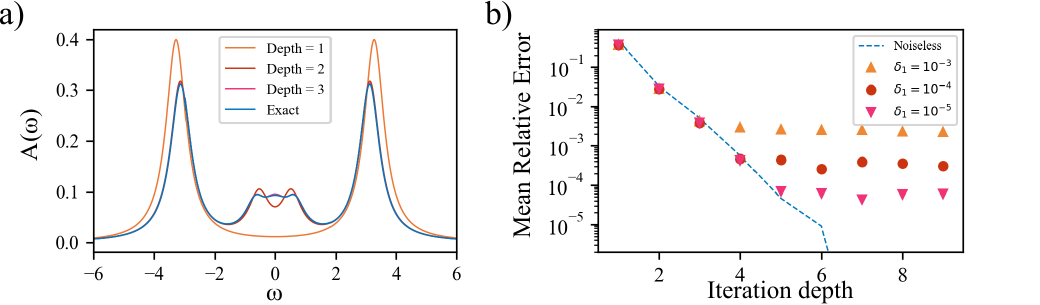}
    \caption{Estimating the Green's function of a SIAM with $4$ bath sites using the ROQAM. 
    In a) we show the spectral function obtained from performing ROQAM with iteration depth $1$, $2$, and $3$, and the exact spectral function. 
    In each case the broadening parameter, $\gamma$, is set to $0.1$ times the Hubbard bandwidth.
    In b) we show the mean relative error versus iteration depth for the ROQAM with and without finite-precision errors.
    We tested the ROQAM with three different levels of initial sampling precision ($\delta_1$) and find that the noisy estimates follow the noiseless error curves until hitting a noise floor where increasing the iteration depth no longer improves the error.
    }
    \label{fig:data_snapshot}
\end{figure*}

For $M = H$, this is equivalent to the Lanczos method.
A quantum computer could access $H$ through a block-encoding, form the orthogonal polynomials using the QSVT, and then estimate the matrix elements in Eq.~\ref{eq:poly_matrix_elements} using a Hadamard test. 
Unfortunately, block-encoding necessarily contains a subnormalization factor $\lambda$ such that the matrix accessed is actually $H/\lambda$ and the attendant polynomials are $P_j(H/\lambda)$.
This would result in the subnormalization factor of the polynomials growing as $\mathcal{O}(\lambda^j)$ and requiring the precision of the Hadamard tests to be similarly scaled.
This is avoided by instead using the time-evolution operator, $M = U(\Delta t)$, and absorbing $\lambda$ into $\Delta t$.
Now the effects of $\lambda$ can be canceled out with linear cost by rescaling $\Delta t \rightarrow \lambda \Delta t$.
Alternatively, $\lambda$ can be sidestepped altogether by approximating $U$ with a product formula.
The Arnoldi iteration now returns the projected matrix $[U]$, which we process into a representation of $H$ as
\begin{equation}
    [H] = i\Delta t^{-1}\log[U].
    \label{eq:H_from_U}
\end{equation}
Substituting $M = U$ into Eq.~\ref{eq:poly_matrix_elements} yields some simplifications.
Namely, each product of polynomials can be factored as
\begin{equation}
    P_{i}^{\dag}(U)UP_{j}(U) = \sum_{l=1}^{j+1} \left(c_{ij}^{(l)}U^l + d_{ij}^{(l)}U^{\dagger l}\right)
    \label{eq:poly_factored}
\end{equation}
where the $c_{ij}^{(l)}$ and $d_{ij}^{(l)}$ coefficients are determined from the previous matrix elements using the recursive definitions of the orthogonal polynomials given in Eq.~\ref{eq:polys}. 
In this manner, the ROQAM iteration is performed by estimating $\bra{\chi_0}U^l\ket{\chi_0}$ for $l = 1, 2, ..., r-1, r$ on a quantum computer (e.g., using amplitude estimation~\cite{Brassard_2002}).
These expectation values are then passed to a classical computer where $[U]$ is built and processed into $[H]$.
In addition to simplifying the data acquisition from the quantum computer, this step enforces that $[U]$ retains its upper-Hessenberg structure even in finite precision, thus improving the method's stability.
The matrix $[H]$ can then be used in Eq.~\ref{eq:A_quad} to estimate $\braket{\chi_0|f(H)|\chi_0}$.
This process is summarized in Fig.~\ref{fig:protocol}.

While the time-evolution operator allows us to avoid the $\mathcal{O}(\lambda^r)$ scaling, there are subtleties to the choice of timestep $\Delta t$, particularly when considering finite sampling errors.
If $\Delta t$ is too large, the matrix logarithm is not single-valued and $[H]$ will not be recovered correctly.
This amounts to the condition
\begin{equation}
    \Delta t \leq \frac{\pi}{|E|_{max}},
    \label{eq:no_aliasing}
\end{equation}
where $|E|_{max}$ is the eigenvalue of $H$ with largest absolute value where the corresponding eigenvector has nonzero overlap with $\ket{\chi_0}$.
If $\Delta t$ is close to zero then $U$ will be close to the identity operator, implying the state $\ket{\Tilde{\chi}^{'}_{j+1}} = U\ket{\chi_j}$ will have high overlap with $\ket{\chi_j}$.
When $\ket{\Tilde{\chi}^{'}_{j+1}}$ is orthogonalized against all previous states, it will then have very low norm, amplifying finite-precision errors upon normalization.
Thus, the optimal choice of $\Delta t$ minimizes this error amplification while satisfying Eq.~\ref{eq:no_aliasing}. 
This seems to be a very difficult condition to satisfy \textit{a priori}, and in practice heuristics are likely needed to select $\Delta t$.

Estimating the outcome of any given circuit to finite precision $\delta$ biases the aggregate estimate in Eq.~\ref{eq:A_quad}.
Exact determination of the bias is instance specific, however, our numerical tests indicated that the bias is close to $\mathcal{O}(\delta)$, as is the case for the Lanczos method~\cite{knizhnerman1996simple, chen2024lanczosalgorithmmatrixfunctions}.
Inspection of Eqs.~\ref{eq:poly_matrix_elements} and \ref{eq:poly_factored}, reveals that as $l$ increases, the expectation value $\braket{\chi_0|U^l|\chi_0}$ appears fewer times in the matrix $[U]$.
This suggests that $\delta_l$, the error to which $\braket{\chi_0|U^l|\chi_0}$ is estimated, can be increased with $l$.
Our tests indicate that $\delta_l$ can be scaled like $\mathcal{O}(l)$.
This results in large computational savings, counteracting the increasing cost of implementing $U^l$.

\textit{Results.--- }We emulate the ROQAM to estimate the diagonal elements of the zero-temperature single-particle GF of the single-impurity Anderson model (SIAM)~\cite{PhysRev.124.41}.
The ground eigenpair $\ket{\psi_0}$ and $E_0$ are inputs to the method, so we begin by explicitly diagonalizing the Hamiltonian, obtaining $\ket{\psi_0}$ and $E_0$.
We then perform two separate Arnoldi iterations to estimate $G^+{(\omega)}$ and $G^-(\omega)$.
For both iterations we use the same $U$, however, we set $\ket{\chi_0} = a^{\dagger}\ket{\psi_0}$ to estimate $G^+{(\omega)}$ and $\ket{\chi_0} = a\ket{\psi_0}$ to estimate $G^-{(\omega)}$.
To emulate the effects of finite-sampling errors, random numbers with variance $\delta_l^2$ are added to the real and imaginary components of each $\braket{\chi_0|U^l|\chi_0}$ calculated.
We set $\delta_l = l\delta_1$, where $\delta_1$ is the error on the first expectation value $\braket{\chi_0|U|\chi_0}$.
For more details on how parameters were chosen, see Appendix~\ref{sec:numbers}.
Once we have estimates of $G^+{(\omega)}$ and $G^-(\omega)$, we add them together to estimate $G{(\omega)}$.

In Fig.~\ref{fig:data_snapshot} we show results for a SIAM at half-filling in its metallic phase with $4$ bath spin-orbitals.
The SIAM is parameterized according to the self-consistent solution of the two-site DMFT problem~\cite{Potthoff_2001} while scaling the hopping terms to maintain a constant hybridization function.
The choice of $\gamma$ greatly affects the rate of convergence, especially at small bath sizes.
If $\gamma$ is too small, the ROQAM is simply being used as an eigensolver to find all eigenvalues within the support of $\ket{\chi_0}$, thus requiring deep iteration and high precision per iterate.
However, if $\gamma$ is too large, key features of the spectral function cannot be resolved.
Ref.~\cite{Lu_2014} suggests setting $\gamma$ to three times the Hubbard bandwidth divided by the number of poles in the Green’s function in order to obtain a continuous spectrum.
In our examples $\gamma$ was simply selected to be one-tenth of the Hubbard bandwidth.
We also show the rate of convergence with respect to iteration depth, both with and without finite-sampling errors.
Finite sampling introduces bias, where the noisy estimates follow the noiseless~\footnote{Here ``noiseless'' means that we did not add random noise to mimic finite-sampling precision. However, all calculations were performed with numpy's double-precision floating-point arithmetic.} error curves until reaching a ``noise floor'', beyond which increasing the iteration depth no longer improves the estimate.

\begin{figure}[!h]
    \vspace{1em}
    \centering
    \includegraphics{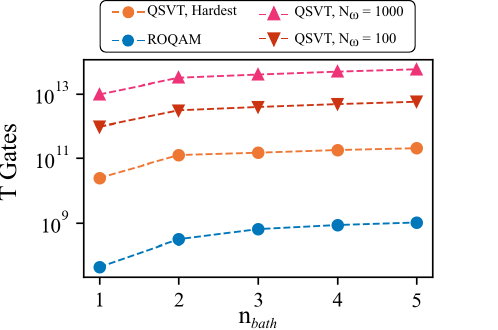}
    \caption{Aggregate T-gate count required to estimate the spectral function of the SIAM with $1$ -- $5$ bath sites to below $1\%$ mean relative error using the ROQAM and QSVT. 
    For QSVT, we report estimates for $\numfreqs = 100$ and $\numfreqs = 1000$ equally spaced frequencies, and the single frequency with the highest $T$ count. 
    The ROQAM is multiple orders of magnitude cheaper than the single hardest QSVT and it produces a converged estimate over the entire frequency interval.}
    \label{fig:tgates}
\end{figure}

In Fig.~\ref{fig:tgates} we estimated the total number of $T$ gates required to achieve better than $1\%$ mean relative error in the spectral function for various bath sizes using both the ROQAM and QSVT matrix inversion.
While the $T$ count is no longer a particularly sharp logical resource estimate~\cite{gidney2024magic}, it is still a crude proxy for the overall cost of implementing an algorithm.
We used a Trotterized time-evolution operation for the ROQAM (see Appendix~\ref{sec:QRE}). 
while for the QSVT method we used a linear combination of unitaries~\cite{LCU} block-encoding with in-line rotations for the \textsc{prepare} oracle.
For QSVT we report the costs of estimating $G(\omega)$ across $\numfreqs=100$ and $\numfreqs=1000$ equally spaced frequencies as well as the single frequency with the highest $T$ count to estimate.
For all instances studied we found that the ROQAM costs roughly two orders of magnitude less than even the single hardest instance of QSVT.
These $T$ counts do not include the cost of preparing $\ket{\chi_0}$ or estimating $E_0$ and therefore should not be taken as end-to-end resource estimates, but rather the costs of these particular subroutines.

This framework can also be extended to estimate GFs at nonzero temperature.
Previous work~\cite{jaklivc1994lanczos, prelovvsek2013ground,gentinetta2026quantumfinitetemperaturelanczos} has combined Krylov methods with stochastic trace estimation~\cite{ubaru2017fast}.
However, this requires building many distinct Krylov subspaces starting from many different states drawn from a Hutchinson distribution~\cite{hutchinson1989stochastic, Shen_2026}.
We show that thermal GFs can be estimated from a single Krylov subspace with the same method as for zero temperature.
The thermofield double (TFD) state represents the thermal state as a pure state with twice as many qubits
\begin{equation}
    \ket{\Psi(\beta)} = \sqrt{\frac{1}{Z(\beta)}}\sum_{n} \left( \mathbb{1} \otimes e^{-\beta H/2} \right) \ket{n}\ket{n}.
\end{equation}
This is a vectorization of the $2^n \times 2^n$ matrix $e^{-\beta H/2}$.
Using the TFD state and properties of vectorization the thermal GF can be written as
\begin{widetext}
\begin{align}
    G_{pq}^+(\beta, \omega) &=  \bra{\Psi(\beta)} \left( \mathbb{1} \otimes a_p \right) \left[ (\omega + i\gamma) + ( H\otimes \mathbb{1} - \mathbb{1} \otimes H)\right]^{-1} \left( \mathbb{1} \otimes a_q^{\dagger} \right) \ket{\Psi(\beta)}  \nonumber,\\
    G_{pq}^-(\beta, \omega) &= \bra{\Psi(\beta)} \left( \mathbb{1} \otimes a_p^{\dagger} \right) \left[ (\omega + i\gamma) - ( H\otimes \mathbb{1} - \mathbb{1} \otimes H)\right]^{-1} \left( \mathbb{1} \otimes a_q \right) \ket{\Psi(\beta)}.   
    \label{eq:G_therm}
\end{align}
\end{widetext}
Using this equation it is straightforward to estimate the thermal GFs with ROQAM by replacing the ground state with the TFD state and finding a representation of $\left( H\otimes \mathbb{1} - \mathbb{1} \otimes H \right)$ instead of $H$ (see Appendix~\ref{sec:toasty}).
This requires no estimates of eigenvalues of $H$.
We note that this representation could also be used in classical subspace methods.

\textit{Conclusion.--- }We have presented a robust quantum Arnoldi method to estimate single-particle GFs.
While this is the most natural application, it is straightforward to extend this method to estimating other matrix functions.
Because the ROQAM is formulated in terms of orthogonal polynomials, the upper-Hessenberg structure of the project matrices is preserved independent of estimation errors.
We have also shown how to extend our method to GFs at nonzero temperature using the TFD formalism.
Ongoing work is aimed at extending the ROQAM at nonzero temperature to applications in high-energy density physics, e.g., calculating linear response functions of warm dense matter~\cite{baczewski2016x,dornheim2023electronic,robinson2026capturing}.

Current limitations of this ROQAM are the lack of tight $\textit{a priori}$ bounds to guide parameter selection (e.g., precision and timestep).
While it is possible that such bounds could be derived in the future this would involve highly nonlinear error propagation, which makes Arnoldi methods difficult to analyze in the presence of noise.
Furthermore, although existing noiseless bounds are given in terms of optimal polynomial approximations, these are notoriously loose and of little predictive power.
For these reasons, we expect that heuristics will be needed for parameter selection.
Finally, as with all iterative methods, a rigorous proof or an extension establishing backwards stability would increase confidence in those heuristics~\footnote{Naturally, one might call this a Stable Orthogonalized Quantum Arnoldi Method (SOQAM).}.

\begin{acknowledgments}
\textit{Acknowledgments.---} 
We thank 
Eric Bobrow,
Daan Camps,
Riley Chien,
Chris Kane,
Alina Kononov,
Andrew Landahl,
Rich Lehoucq,
Ryan Levy,
Cole Maurer,
Mason Rhodes,
and Andrew Zhao
for helpful technical discussions.
This work was supported by the DOE Office of Fusion Energy Sciences ``Foundations for quantum simulation of warm dense matter'' project.

This article has been co-authored by employees of National Technology \& Engineering Solutions of Sandia, LLC under Contract No. DE-NA0003525 with the U.S. Department of Energy (DOE). The authors own all right, title and interest in and to the article and are solely responsible for its contents. The United States Government retains and the publisher, by accepting the article for publication, acknowledges that the United States Government retains a non-exclusive, paid-up, irrevocable, world-wide license to publish or reproduce the published form of this article or allow others to do so, for United States Government purposes. The DOE will provide public access to these results of federally sponsored research in accordance with the DOE Public Access Plan \url{https://www.energy.gov/downloads/doe-public-access-plan}.

\end{acknowledgments}

\bibliography{references}

\clearpage
\widetext
\begin{center}
\textbf{\large Supplemental Materials: \papertitle}
\end{center}

\setcounter{section}{0}
\setcounter{page}{1}
\setcounter{secnumdepth}{2}
\appendix
\makeatletter

These Supplemental Materials elaborate on the central results in the main manuscript.
\begin{itemize}
    \item Appendix~\ref{sec:Arn} reviews the Arnoldi algorithm for evaluating matrix functions.
    \item Appendix~\ref{sec:qArn} describes our quantum adaptation of the Arnoldi algorithm, including error analyses.
    \item Appendix~\ref{sec:green} reviews the single-particle Green's function and derives a spectral form of the thermal Green's function, allowing straightforward extension of Arnoldi methods to nonzero temperature.
    \item Appendix~\ref{sec:QSVT} shows how QSVT could be used to estimate the single-particle Green's function.
    \item Appendix~\ref{sec:numbers} describes the numerical simulations, the effect of finite precision, and how the resource estimates were obtained.
\end{itemize}

\section{The Arnoldi method}
\label{sec:Arn}

In this Appendix we describe the Arnoldi method for evaluating matrix functions~\cite{arnoldi}.
We first define Krylov subspaces and a few of their relevant properties. 
We then discuss how to project a matrix into a Krylov subspace through Arnoldi iteration.
Finally, we describe how to use this projection to estimate expectation values of matrix functions with Arnoldi quadrature.

\subsection{Krylov subspaces}

A Krylov subspace is the span of the vectors formed by repeatedly multiplying a matrix with some fiducial initial vector.
Namely the $r$th Krylov subspace generated by matrix $M$ and vector $\ket{\chi_0}$ is defined as
\begin{equation}
    \mathcal{K}_{r}(M,\ket{\chi_0}) = \mathrm{span}\{\ket{\chi_0}, M\ket{\chi_0}, M^2\ket{\chi_0}, ... M^{r-1}\ket{\chi_0} \}.
\end{equation}
While this is the span of $r$ vectors, the dimension of this subspace can be less than $r$ because these vectors are not guaranteed to be linearly dependent.
There is a critical value $r_0$ beyond which further multiplications by $M$ no longer produce linearly independent vectors.
At $r = r_0$ the Krylov subspace is stabilized such that
\begin{equation}
    \forall j > 0 \hspace{1em} \mathcal{K}_{r_0 + j}(M,\ket{\chi_0}) = \mathcal{K}_{r_0}(M,\ket{\chi_0}).
\end{equation}

The value of $r_0$ is determined by the spectral support of $\ket{\chi_0}$ on $M$.
That is, if $\ket{\chi_0}$ is supported on $k$ eigenvectors of $M$, then
\begin{equation}
    \dim \left( \mathcal{K}_{r}(M,\ket{\chi_0}) \right) = \min(k,r).
    \label{eq:K_dim}
\end{equation}
For example, if $\ket{\chi_0}$ is an eigenvector of $M$ with eigenvalue $\mu$, then $r_0 = 1$ because $M^{j}\ket{\chi_0} = \mu^j\ket{\chi_0}$.

The condition in Eq.~\ref{eq:K_dim} reveals a connection between the Krylov subspaces generated by $M$ and functions of $M$, $f(M)$.
Because $M$ and $f(M)$ share an eigenspace, $\ket{\chi_0}$ has the same spectral support on $M$ and $f(M)$.
This implies that that the Krylov subspaces $\mathcal{K}_{r}(M,\ket{\chi_0})$ and $\mathcal{K}_{q}(f(M),\ket{\chi_0})$ stabilize in the same number of steps.
This also implies that these subspaces stabilize to the same subspace,
\begin{equation}
    \mathcal{K}_{r_0}(f(M),\ket{\chi_0}) = \mathcal{K}_{r_0}(M,\ket{\chi_0}).
    \label{eq:stabilized_subspaces}
\end{equation}

Krylov subspaces provide a natural setting for estimating expectation values of matrix functions $\bra{\chi_0}f(M)\ket{\chi_0}$ because they contain all polynomials of $M$, with degree less than $r$, acting on $\ket{\chi_0}$
\begin{equation}
    \forall p(M) \mathrel{|} \mathrm{deg}(p) < r, \hspace{1em} p(M)\ket{\chi_0} \in \mathcal{K}_{r}(M,\ket{\chi_0}).
    \label{eq:K_polys}
\end{equation}
Furthermore, computations on this subspace are significantly cheaper than on the original space because the dimension of $\mathcal{K}_r(M, \ket{\chi_0})$ is much lower than the dimension of the original space.

\subsection{Arnoldi iteration}

The Arnoldi method takes advantage of the information contained in $\mathcal{K}_{r}(M,\ket{\chi_0})$ by projecting $M$ into this subspace, then evaluating $f(\cdot)$ on the projection.
This projection is performed by constructing an orthonormal basis from $\mathcal{K}_{r}(M,\ket{\chi_0})$ and then evaluating the action of $M$ on this basis.
The process by which $M$ is projected into $\mathcal{K}_{r}(M,\ket{\chi_0})$ is known as Arnoldi iteration~\cite{arnoldi}.

\subsubsection{Standard formulation}
\label{sec:standard_form}

Perhaps the simplest way to obtain an orthonormal basis from $\mathcal{K}_{r}(M,\ket{\chi_0})$ would be to first build $\mathcal{K}_{r}(M,\ket{\chi_0})$ and then apply Gram-Schmidt orthogonalization.
However, this process is numerically unstable, particularly so when the vectors in $\mathcal{K}_{r}(M,\ket{\chi_0})$ are nearly linearly dependent~\cite{chen2024lanczosalgorithmmatrixfunctions}.
Instead, the Arnoldi iteration alternates between expanding the subspace and orthogonalizing.
Starting with an initial vector $\ket{\chi_0}$, we define
\begin{equation}
    \ket{\Tilde{\chi}^{'}_{j+1}} = M\ket{\chi_j}.
\end{equation}
We then orthogonalize against all previous vectors
\begin{equation}
    \ket{\Tilde{\chi}_{j+1}} = \ket{\Tilde{\chi}^{'}_{j+1}} - \sum_{i=0}^{j} \braket{\chi_i|\Tilde{\chi}^{'}_{j+1}}\ket{\chi_i},
    \label{eq:orth_step}
\end{equation}
and normalize to obtain the $(j+1)$th basis state
\begin{equation}
    \ket{\chi_{j+1}} = \frac{\ket{\Tilde{\chi}_{j+1}} }{|| \ket{\Tilde{\chi}_{j+1}} ||}.
    \label{eq:normal_sm}
\end{equation}
Note that if the Krylov subspace stabilizes at step $j$, the orthogonalization step will annihilate the $(j+1)$th vector.
That is, if $j = r_0$, then
\begin{equation}
    \ket{\Tilde{\chi}_{r_0+1}} = \ket{\Tilde{\chi}^{'}_{r_0+1}} - \sum_{i=0}^{j} \braket{\chi_i|\Tilde{\chi}^{'}_{r_0+1}}\ket{\chi_i} = 0.
\end{equation}
This is referred to as the ``breakdown condition'', heralding that the iteration can be terminated.
The Arnoldi iteration is carried out until $j+1 = r$ or breakdown occurs, creating an orthonormal basis which we refer to as the Arnoldi basis.

During the iteration, we simultaneously build $[M]$, the representation of $M$ in the Arnoldi basis, by evaluating the matrix elements
\begin{equation}
    [M]_{ij} = \bra{\chi_i}M\ket{\chi_j}.
\end{equation}
Given the construction of the Arnoldi basis we make some simplifications.
Since $M\ket{\chi_{j}} = \ket{\Tilde{\chi}_{j+1}^{'}}$, we write
\begin{equation}
    [M]_{ij} = \braket{\chi_i|\Tilde{\chi}_{j+1}^{'}}.
\end{equation}
The matrix elements vanish when $i > j+1$, since $\ket{\chi_i}$ is orthogonalized against $\ket{\Tilde{\chi}_{j+1}^{'}}$ during the iteration,
\begin{equation}
    [M]_{ij} = 0 \hspace{1em} \text{if} \hspace{0.5em} i > j+1.
\end{equation}
When $i = j+1$, 
\begin{align}
    \braket{\chi_{j+1}|\Tilde{\chi}_{j+1}^{'}} &= \bra{\chi_{j+1}} \left( \ket{\Tilde{\chi}_{j+1}} + \sum_{i=0}^{j}\braket{\chi_i|\Tilde{\chi}_{j}^{'}}\ket{\chi_{i}} \right) \nonumber \\
    &= \braket{\chi_{j+1}|\Tilde{\chi}_{j+1}} \nonumber \\
    &= || \ket{\Tilde{\chi}_{j+1}} ||.
\end{align}
Thus, the elements of $[M]$ are 
\begin{equation}
    [M]_{ij} = \begin{cases}
                \bra{\chi_i}M\ket{\chi_j}, & \text{if } i \leq j \\
                || \ket{\Tilde{\chi}_{j}} ||, & \text{if } i = j + 1 \\ 
                0, & \text{if } i > j + 1, \\
             \end{cases}
    \label{eq:matrix_elements}
\end{equation}
i.e., $[M]$ is upper Hessenberg (its matrix elements are zero below the first sub-diagonal).
For this reason, the projected matrix is often denoted as $H$.
Because it is common to use $H$ to denote a Hamiltonian, we instead use square brackets $[\cdot]$ to distinguish between the matrix $M$ and its representation in the Krylov subspace $[M]$.

Equivalently, one can define $Q$ as the projection matrix from the original vector space into the Arnoldi basis
\begin{equation}
    Q =  \begin{bmatrix} 
            \ket{\chi_0} & \ket{\chi_1} & ... & \ket{\chi_{r-1}}
        \end{bmatrix},
\end{equation}
such that $[M]$ is simply the projection of $M$ into the Arnoldi basis
\begin{equation}
    [M] = Q^{\dagger}MQ.
\end{equation}

\subsubsection{Orthogonal polynomial formulation}
\label{sec:poly_form}

The Arnoldi iteration can also be formulated in terms of recursively defined orthogonal polynomials.
We define $p_{j+1}(M)$ as the degree-$(j+1)$ polynomial of $M$ that generates the $(j+1)$th Arnoldi state when applied to $\ket{\chi_0}$,
\begin{equation}
    p_{j+1}(M)\ket{\chi_0} = \ket{\chi_{j+1}}.
\end{equation}
These polynomials satisfy the orthogonality condition
\begin{equation}
    \bra{\chi_0}p^{\dagger}_{i}(M)p_j(M)\ket{\chi_0} = \delta_{ij}.
\end{equation}
Finding these polynomials is then equivalent to constructing the Arnoldi basis.

These polynomials are recursively defined as
\begin{subequations}
\begin{align}
    p_0(M) &= 1 \\
    \tilde{p}_{j+1}(M) &= (M- [M]_{jj})p_{j}(M) - \sum_{i=0}^{j-1}[M]_{ij}p_i(M).
    \label{eq:polys_sm}
\end{align}
\end{subequations}
Here, the tilde denotes the fact that the associated Arnoldi vector, $\ket{\Tilde{\chi}_{j+1}}$, is not normalized.
The norm of $\ket{\Tilde{\chi}_{j+1}}$ is then
\begin{equation}
    || \ket{\Tilde{\chi}_{j+1}} || = \sqrt{ \braket{\chi_0|\Tilde{p}^{\dag}_{j+1}(M)\Tilde{p}_{j+1}(M)|\chi_0 } },  
\end{equation}
and thus the properly normalized polynomial is 
\begin{equation}
    p_{j+1}(M) = \frac{\Tilde{p}_{j+1}(M)}{|| \ket{\Tilde{\chi_{j+1}}} ||}.
\end{equation}
Following Equation~\ref{eq:matrix_elements}, the matrix $[M]$ is defined in terms of these polynomials:
\begin{equation}
    [M]_{ij} = \begin{cases}
                \bra{\chi_0}p^{\dag}(M)Mp(M)\ket{\chi_0}, & \text{if } i \leq j \\[0.5em]
                \sqrt{ \braket{\chi_0|\Tilde{p}^{\dag}_j(M)\Tilde{p}_j(M)|\chi_0 } }, & \text{if } i = j + 1 \\[0.5em] 
                0, & \text{if } i > j + 1. \\
             \end{cases}
    \label{eq:poly_matrix_elements_sm}
\end{equation}

\subsection{Arnoldi quadrature}

After $[M]$ is found, we estimate the action of some matrix function $f(M)$ on the starting vector $\ket{\chi_0}$ by evaluating $f([M])$.
It is considerably cheaper to evaluate $f([M])$ than $f(M)$, because the dimension of $[M]$ is much smaller than that of $M$.
Formally, we approximate the quadratic form of $f(M)$ as
\begin{equation}
    \braket{\chi_0|f(M)|\chi_0} \approx \braket{\chi_0|Q^{\dag}f([M])Q|\chi_0}.
\end{equation}
This expression might seem to imply that we are explicitly storing $Q$ and projecting $f([M])$ back to the higher-dimensional Hilbert space where $\ket{\chi_0}$ lives.
This is not the case, however, as $\braket{\chi_0|Q^{\dag}f([M])Q|\chi_0}$ is simply the top-left element of the matrix $f([M])$ multiplied by the norm squared of $\ket{\chi_0}$.
So, we can also express our approximation as
\begin{equation}
    \braket{\chi_0|f(M)|\chi_0} \approx \braket{\chi_0|\chi_0}f([M])_{00},
    \label{eq:A_quad_sm}
\end{equation}
and we see that we can approximate $\braket{\chi_0|f(M)|\chi_0}$ without explicitly storing $Q$. 
Thus, once we have obtained the $r \times r$ matrix $[M]$, we work only with that matrix.

In some cases, there are error bounds that can be applied to the approximation in Eq.~\ref{eq:A_quad_sm}.
We first consider cases in which there are no finite-precision errors. 
If breakdown occurs then the Krylov subspace is stabilized and thus invariant under multiplication by $M$.
In this case, the action of $M$ on $\ket{\chi_0}$ is exactly captured by $[M]$ and the estimate in Eq.~\ref{eq:A_quad_sm} is exact.
When $r < r_0$ and $M = H$ is a Hermitian matrix, the Arnoldi process is equivalent to the Lanczos method~\cite{lanczos} and the approximation error is bounded by the error of an optimal degree-$2r$ polynomial approximation of $f(H)$ on the real axis~\cite{gautschi2004orthogonal, golub2009matrices},
\begin{equation}
    \left| \braket{\chi_0|f(H)|\chi_0} - \braket{\chi_0|\chi_0}f([H])_{00} \right| \leq 2||\ket{\chi_0}||^2 \min_{\textrm{deg}(p)<2r} ||f - p||.
\end{equation}
Similar bounds have also been derived for Arnoldi quadrature when the target matrix is unitary.
Complications arise in this scenario because the representation $[U]$ will not generally be unitary.
This is dealt with by projecting $[U]$ to the nearest unitary.
The resulting error is then
\begin{equation}
    \left| \braket{\chi_0|f(U)|\chi_0} - \braket{\chi_0|\chi_0}f([U])_{00} \right| \leq 2||\ket{\chi_0}||^2 \min_{p \in \mathbb{L}_{r-1}} ||f - p||, 
\end{equation}
where $\mathbb{L}_{r-1}$ is the set of Laurent polynomials of degree $r-1$~\cite{kirby2025quantumkrylovalgorithmszego}.
These bounds in terms of optimal polynomial approximations are straightforward, because the Krylov subspace contains these polynomials by construction (see e.g., Eq.~\ref{eq:K_polys}).

Good \textit{a priori} bounds are much more difficult to derive in the presence of finite-precision errors.
For Hermitian $H$, in which all matrix-vector calculations are performed to precision $\delta$, it has been shown that convergence slows down by at most an $\mathcal{O}(r)$ multiplicative factor, leading the approximation to become biased by at most an $\mathcal{O}(\text{poly}(r)\delta)$ factor~\cite{knizhnerman1996simple}
\begin{equation}
    \left| \braket{\chi_0|f(H)|\chi_0} - \braket{\chi_0|\chi_0}f([H])_{00} \right| \leq \mathcal{O}(r)||\ket{\chi_0}||^2 \min_{\textrm{deg}(p)<2r} ||f - p|| + \mathcal{O}(||f||\mathrm{poly}(r)\delta).
    \label{eq:noisy_bound}
\end{equation}

It should be noted that all of these bounds are very loose in practice and typically do not well predict actual performance~\cite{chen2024lanczosalgorithmmatrixfunctions}.
Even in finite precision, the estimate in Eq.~\ref{eq:A_quad_sm} tends to outperform the bounds that rely on exact arithmetic.

\section{The robust quantum Arnoldi method}
\label{sec:qArn}

In the previous Appendix we reviewed the Arnoldi method applied to general matrices.
We now specialize to the case where $M$ is a Hamiltonian $H$, as is the case in evaluating Green's functions.
We have seen that the Arnoldi method for matrix-function estimation is upper bounded by optimal polynomial approximations and even tends to outperform these bounds.
Furthermore, once the projected matrix $[H]$ is found, it can be reused for as many functions as one wishes to evaluate.
The primary bottleneck is the Arnoldi iteration where the basis states must be explicitly stored and manipulated.
This motivates the ROQAM where the iteration is performed using a quantum computer.

By storing computation information in its wave function, quantum computers may provide an exponential advantage in the storage and manipulation of the Arnoldi basis states.
However, there are particular traits we must take into account when designing a quantum algorithm to perform  the Arnoldi iteration.
Of primary concern is the fact that the quantities we calculate during the iteration will be estimated at finite precision.
Even estimation at the Heisenberg limit (e.g., using quantum amplitude estimation~\cite{Brassard_2002}) still has $\mathcal{O}(1/\epsilon)$ overhead, compared with $\text{polylog}(1/\epsilon)$ arithmetic on classical computers.
Thus, the effects of finite-precision errors are more salient.

\subsection{Inputs}

In order to perform the Arnoldi iteration on a quantum computer we require access to $H$ and the initial vector $\ket{\chi_0}$.
We assume efficient access to a $(\lambda, m, \epsilon)$ block-encoding~\cite{Low_2019, Chakraborty_2019, lin2022lecturenotesquantumalgorithms} of $H$ defined as
\begin{equation}
    \left\|(\bra{0^m} \otimes I)U_M(\ket{0^m} \otimes I) - \frac{H}{\lambda} \right\| \leq \epsilon.
\end{equation}
Efficient initialization of $\ket{\chi_0}$ is a more stringent requirement that generally cannot be guaranteed.
State preparation routines are the subject of intensive research~\cite{Lin_2020, chen2023quantum, huggins2025efficient, Zhan_2026} and outside the scope of this work.

While errors in the block-encoding of $H$ and preparation of $\ket{\chi_0}$ effect the estimates produced by the Arnoldi method, we argue they do not do so in a unique way.
This can be seen as follows: assume that instead of $\ket{\chi_0}$ we instead begin with a noisy state
\begin{equation}
    \ket{\xi} = \sqrt{p}\ket{\chi_0} + \sqrt{1-p}\ket{\chi_{\perp}}.
\end{equation}
Further assume that instead of a block-encoding of $H$, we instead access $H'$.
In this case the Arnoldi method will estimate $\bra{\xi}f(H')\ket{\xi}$ instead of $\bra{\chi_0}f(H)\ket{\chi_0}$.
If we still use this as our estimate of $\bra{\chi_0}f(H)\ket{\chi_0}$ our estimate will be biased as
\begin{equation}
    \eta = \bra{\chi_0}f(H)\ket{\chi_0} - \bra{\xi}f(H')\ket{\xi}.
\end{equation}
While this bias is clearly undesirable, it is not unique to the Arnoldi method and would be there if another method, such as the quantum singular value transformation (QSVT)~\cite{Gily_n_2019}, was used to estimate $f(H)$ given noisy inputs $\ket{\xi}$ and $H'$.
If errors on these inputs are high enough, it may impact the convergence rate of the Arnoldi method.
However, in this case, we are simply setting up an entirely different problem than intended.

For the remainder of this Appendix, we assume that there are no errors in these inputs.
We will not make any assumptions on the block-encoding subnormalization factor $\lambda$.
As we will see in the following section, $\lambda$ plays an important role in performing the Arnoldi iteration on a quantum computer.

\subsection{Quantum Arnoldi iteration}
\label{sec:QAI}

While we could perform the Arnoldi iteration on a quantum computer using the standard formulation in Appendix~\ref{sec:standard_form}, the orthogonal polynomial formulation in Appendix~\ref{sec:poly_form} is more efficient in both space and time.
The standard formulation requires storage of the entire Arnoldi basis, which would require $r$ copies of the system register for $r$ levels of iteration.
Furthermore, performing the orthogonalization across these system registers requires complicated quantum circuits~\cite{Zhang_2021}.
In the orthogonal polynomial formulation, however, the orthogonalization is performed implicitly while only requiring a single system register, leading to much simpler quantum circuits.

Given access to a block-encoding of $H$, we could use the QSVT to form the orthogonal polynomials and then use Hadamard tests to estimate the matrix elements of $[H]$ (Equation~\ref{eq:poly_matrix_elements_sm}).
Since $H$ is Hermitian, this would reduce the Arnoldi method to the special case of the Lanczos method.
However, we must remember that our block-encoding actually gives us access to $H/\lambda$, where $\lambda$ is lower bounded by the spectral norm of $H$.
As such, QSVT constructs polynomials $P_j(H/\lambda)$, causing the subnormalization factor to grow like $\mathcal{O}(\lambda^j)$.
To counteract this, the precision of the Hadamard tests would need to be similarly scaled, resulting in an overall complexity of $\mathcal{O}(\lambda^r)$ to perform $r$ levels of iteration.

Instead of $H$ directly, we use $U(H) = e^{-i(H/\lambda) \Delta t}$.
Now, the $1/\lambda$ factor is absorbed into $\Delta t$ and is counteracted with linear $\mathcal{O}(\lambda)$ cost by rescaling $\Delta t \rightarrow \lambda\Delta t$.
Alternatively, the issue with $\lambda$ can be sidestepped altogether through a Trotter product construction of $U$. 
We perform the Arnoldi iteration on $U$ obtaining the representation $[U]$, which we post-processes into a representation of $[H]$ as
\begin{equation}
    [H] = i\Delta t^{-1}\log[U].
    \label{eq:H_from_U}
\end{equation}

Substituting $M = U$ into Equation~\ref{eq:poly_matrix_elements_sm} yields some simplifications.
Namely, every polynomial can be factored into sums of autocorrelation functions
\begin{equation}
    \bra{\chi_0}P_{i}^{\dag}(U)UP_{j}(U)\ket{\chi_0} = \sum_{l=1}^{j+1}c_l^{ij}\bra{\chi_0}U^l\ket{\chi_0} + \sum_{l=1}^{j+1}d_l^{ij}\bra{\chi_0}U^l\ket{\chi_0}^*
    \label{eq:auto_corrs}
\end{equation}
where the $c_l^{ij}$ and $d_l^{ij}$ coefficients are determined by the previous matrix elements using the recursive definitions of the orthogonal polynomials given in Eq.~\ref{eq:polys}. 
In this manner, the quantum Arnoldi iteration is performed by estimating $\bra{\chi_0}U^l\ket{\chi_0}$ for $l = 1, 2, ..., r-1, r$ on a quantum computer.
These expectation values are then passed to classical computer where $[U]$ is built and processed into $[H]$.
The matrix $[H]$ is now ready to be used in Equation~\ref{eq:A_quad} to estimate $\braket{\chi_0|f(H)|\chi_0}$.

We note that from Eq.~\ref{eq:stabilized_subspaces} we know that the Krylov subspaces generated by $H$ and $U(H)$ always stabilize to the same subspace in the same number of steps.
This implies that, in a certain sense, using $U(H)$ for the Arnoldi iteration is as good as $H$ when the matrix function $f(\cdot)$ requires the Krylov subspace to have stabilized for the approximation to be exact.

\subsection{Choice of timestep}
\label{sec:delta_t}

In switching from $H/\lambda$ to $U = e^{-iH/\lambda \Delta t}$ we have introduced a new parameter $\Delta t$.
The benefit is that $\Delta t$ can be tuned to cancel out adverse effects of $\lambda$, however, the choice of $\Delta t$ greatly affects the performance of the method in the presence of finite sampling precision.
Before considering the effects of finite precision we first show restrictions on $\Delta t$ even in infinite precision.

In estimating the expectation values $\braket{\chi_0|U^l|\chi_0}$ we are sampling from the signal $\braket{\chi_0|U(t)|\chi_0}$ with sampling frequency
\begin{equation}
    \omega_s = \frac{2\pi}{\Delta t}.
\end{equation}
In order to avoid aliasing, the sampling frequency must be at least twice as large as the largest frequency component in the signal.
Thus we must set
\begin{equation}
    \Delta t \leq \frac{\pi}{|E|_{\textrm{max}}}.
    \label{eq:no_aliasing_sm}
\end{equation}
Where $|E|_{\textrm{max}}$ is the eigenvalue with largest absolute value where the corresponding eigenvector has nonzero overlap with $\ket{\chi_0}$.
While $|E|_{\textrm{max}}$ will not generally be known, it is bounded by the spectral norm of $H$
\begin{equation}
    |E|_{\textrm{max}} \leq ||H||_2
\end{equation}
meaning we could set $\Delta t$ as
\begin{equation}
    \Delta t \leq \frac{\pi}{||H||_2},
\end{equation}
however, this may be overly restrictive when considering finite-precision effects.

In addition to this maximum set by the no-aliasing condition, we obviously cannot let $\Delta t = 0$.
At $\Delta t = 0$, $U = \mathbb{1}$ and carries no information about $H$.
Equivalently, since $\ket{\chi_0}$ is trivially an eigenstate of $\mathbb{1}$, the norm of $\ket{\Tilde{\chi}_1}$ is $0$.

This leads naturally to consideration of the effects of finite precision.
If $\Delta t$ is near zero the norms of the Arnoldi states will be small, amplifying the finite-precision errors during the normalization step.
Thus in the presence of noise $\Delta t$ cannot be too close to zero.
Actually, it is not so simple, since the requirement is really that the norms of the Arnoldi states not be too small.
While the norms will always be small when $\Delta t$ is near zero, the norms have more complicated sinusoidal dependence on $\Delta t$.
For example, if we express $\ket{\chi_0}$ in the spectral basis of $H$
\begin{equation}
    \ket{\chi_0} = \sum_n c_n\ket{n}
\end{equation}
then the norm of the first state is
\begin{equation}
    |\braket{\Tilde{\chi}_1|\Tilde{\chi}_1}| = \sqrt{1 - \sum_n \left( |c_n|^4 \right) - \sum_{m>n} |c_m|^2|c_n|^2\cos{(E_m - E_n)\Delta t}  }
\end{equation}
With the $j > 1$ states having ever more complicated expressions.
The norms should be maximized in order to minimize the forward propagation of error.
This appears to be a very difficult task, even for $\ket{\Tilde{\chi}_1}$ this requires knowledge of the spectrum of $H$ and the spectral decomposition of $\ket{\chi_0}$.

One may wonder if the qubitized walk operator~\cite{Low_2019}, $W = \exp\left(i\arccos\left(H/\lambda\right)\right)$, may be used as the generating matrix instead of the time evolution operator.
This is not possible, since $W$ does not have a means to cancel out $\lambda$.
When $\lambda$ is large, $W$ is close to the identity, meaning after orthogonalization the Arnoldi states will have very low norm.

\subsection{Estimation precision}
\label{sec:precision}

As we have seen from Appendix~\ref{sec:QAI}, the quantum Arnoldi iteration reduces to the estimation of $\braket{\chi_0|U^l|\chi_0}$ for $l = 0, 1, ..., r$.
We see from Eq.~\ref{eq:noisy_bound}, and from simulated data, that the main effect of finite estimation precision is to introduce a bias proportional to the precision.
Tight \textit{a priori} bounds are difficult to derive in the presence of finite precision.
In lieu of these bounds, heuristics can be used to set the estimation precision.

Perhaps the simplest strategy would be to select a $\delta$ on the order of the desired error on the estimate of $\braket{\chi_0|f(H)|\chi_0}$, and then estimate every $\braket{\chi_0|U^l|\chi_0}$ to error $\delta$.
However, as we see from Equation~\ref{eq:auto_corrs}, not every estimate $\braket{\chi_0|U^l|\chi_0}$ is used the same number of times.
This suggests a hierarchy where as $l$ increases, the error to $\braket{\chi_0|U^l|\chi_0}$ is estimated can be increased
\begin{equation}
    \delta_0 < \delta_1 < ... < \delta_r.
\end{equation}
This allows for significant savings since it can partially counteract the difficulty of applying $U^l$ for increasing $l$.
Exact error propagation from each estimate is difficult and once again requires \textit{a priori} knowledge of the norms of the Arnoldi states, motivating heuristics to set an error budget across iterates.
We defer a discussion of the heuristics we used to Appendix~\ref{sec:numbers}, where we discuss the numerical tests we conducted.

\section{The single-particle Green's function}
\label{sec:green}

The ROQAM can be used to estimate single-particle Green's functions both at zero and nonzero temperatures using the spectral representation of the Green's function.
First, we define the forwards and backwards Green's functions in the time domain assuming a time-independent Hamiltonian as

\begin{subequations}
\begin{align}
    G_{pq}^+(t) &= \Theta(t)\braket{U^{\dag}(t)a_{p}U(t)a_{q}^{\dag}}~\text{and}\\
    G_{pq}^-(t) &= \Theta(-t)\braket{U^{\dag}(t)a_{p}^{\dag}U(t)a_{q}}.
\end{align}
\end{subequations}

\subsection{Zero-temperature Green's functions}

At zero temperature these expectation values are evaluated with respect to the ground state
\begin{subequations}
\begin{align}
    G_{pq}^+(t) &= \Theta(t)\bra{\psi_0}U^{\dag}(t)a_{p}U(t)a_{q}^{\dag}\ket{\psi_0}~\text{and} \\
    G_{pq}^-(t) &= \Theta(-t)\bra{\psi_0}U^{\dag}(t)a_{p}^{\dag}U(t)a_{q}\ket{\psi_0}.
\end{align}
\end{subequations}
The Green's functions are expressed in the frequency domain through a Fourier transform, however, they will contain poles at the eigenvalues of $H$ and thus the Fourier transforms are not well defined everywhere along the real axis.
To handle this, a broadening parameter ($i\gamma$) is added to the frequency and the Green's function on the real axis is defined in the limit that $\gamma \rightarrow 0_+$.
Substituting $U(t) = e^{-iHt}$, the forward Green's function is given as
\begin{align}
    G_{pq}^+(\omega) &= \int_{-\infty}^{\infty}dt \Theta(t)\bra{\psi_0}e^{iHt}a_{p}e^{-iHt}a_{q}^{\dag}\ket{\psi_0}e^{i(\omega + i\gamma)t} \nonumber \\
    &= \bra{\psi_0}a_{p}\int_{0}^{\infty}dte^{-i (\omega + i\gamma - (H - E_0))t}a_{q}^{\dag}\ket{\psi_0} \nonumber \\
    &= \bra{\psi_0}a_{p}\left[ (\omega+i\gamma) - (H - E_0) \right]^{-1}a_{q}^{\dag}\ket{\psi_0},
\end{align}
where we have used the definition of the resolvent operator
\begin{equation}
    \int_{0}^{\infty}dte^{-i (x - A)t} = R(x - A) = \left[ x - A \right]^{-1}.
\end{equation}
The backwards Green's function is similarly defined, thus we have
\begin{subequations}
\begin{align}
     G^+_{pq}(\omega) &= \bra{\psi_0}a_{p}\left[ (\omega+i\gamma) - (H - E_0) \right]^{-1}a_{q}^{\dag}\ket{\psi_0}~\text{and} \\
     G^-_{pq}(\omega) &= \bra{\psi_0}a_{p}^{\dag}\left[ (\omega+i\gamma) + (H - E_0) \right]^{-1}a_{q}\ket{\psi_0}.
     \label{eq:G(w)_sm}
\end{align}
\end{subequations}

It is straightforward to obtain diagonal ($p = q$) elements of $G$ from this representation.
For $G_{pp}^+(\omega)$, set $\ket{\chi_0} = a_p^{\dag}\ket{\psi_0}$ and perform the Arnoldi iteration using $e^{-iH\Delta t}$.
We then estimate the Green's function as
\begin{align}
    G_{pp}^+(\omega) & \approx \left[ \left( (\omega +i\gamma + E_0)\mathbb{1}_r - [H] \right)^{-1} \right]_{00}.
\end{align}
Once estimates of $G_{pp}^+(\omega)$ and $G_{qq}^+(\omega)$ are obtained, we can estimate the off diagonals $G_{pq}^+(\omega)$ and $G_{qp}^+(\omega)$ using two more iterations.
Namely we perform two more instances of Arnoldi iteration, once with $\left(a_p^{\dag} + a_q^{\dag}\right)\ket{\psi_0}$ and the other with $\left(a_p^{\dag} + ia_q^{\dag}\right)\ket{\psi_0}$ as the starting states.
The off diagonals are then given as
\begin{align}
    G_{pq}^+(\omega) & \approx \frac{1}{2}\left( \left[\left( (\omega +i\gamma + E_0)\mathbb{1}_r - [H^{+}]  \right)^{-1} \right]_{00} - G_{pp}^+(\omega) - G_{qq}^+(\omega) \right) \nonumber \\
    &- \frac{i}{2}\left( \left[ \left( (\omega + i\gamma + E_0)\mathbb{1}_r - [H^{+i}] \right)^{-1} \right]_{00} - G_{pp}^+(\omega) - G_{qq}^+(\omega) \right) \\
    G_{pq}^+(\omega) & \approx \frac{1}{2}\left( \left[ \left( (\omega + i\gamma + E_0)\mathbb{1}_r - [H^{+}] \right)^{-1} \right]_{00} - G_{pp}^+(\omega) - G_{qq}^+(\omega) \right) \nonumber \\
    &+ \frac{i}{2}\left( \left[ \left( (\omega + i\gamma + E_0)\mathbb{1}_r - [H^{+i}] \right)^{-1} \right]_{00} - G_{pp}^+(\omega) - G_{qq}^+(\omega) \right).
\end{align}
Where $[H^+]$ is the representation obtained starting from $\left(a_p^{\dag} + a_q^{\dag}\right)\ket{\psi_0}$ and $[H^{i+}]$ is the representation obtained starting from $\left(a_p^{\dag} + ia_q^{\dag}\right)\ket{\psi_0}$
The backwards Green's functions are found in exactly the same way, the only difference is that the starting states use the annihilation instead of creation operators (e.g., $a_p\ket{\psi_0}$).

Once the representations of $H$ are found, the Green's functions are estimated for as many frequencies as desired.

\subsection{Thermal Green's functions}
\label{sec:toasty}

It might seem more complicated to apply the Arnoldi method at nonzero temperature.
The main difficulty lies in expressing the Green's function in the frequency domain when the initial state is not an eigenstate of $H$, as is the case for $T = 0$.
In obtaining Equation~\ref{eq:G(w)_sm} we used the fact that $\bra{\psi_0}e^{iHt} = \bra{\psi_0}e^{iE_0t}$ in order to commute the time-evolution operator past the $a$ or $a^{\dag}$. 
However, at $T > 0$, the Green's function is now obtained as 
\begin{equation}
    G_{pq}^+(\beta, t) = \Theta(t)\frac{\mathrm{Tr}(e^{-\beta H/2}e^{iHt}a_pe^{-iHt}a_q^{\dag}e^{-\beta H/2})}{Z(\beta)},
    \label{eq:therm_G(t)}
\end{equation}
where $Z(\beta)$ is the partition function at inverse temperature $\beta$.
The trace can be expanded as
\begin{equation}
    \mathrm{Tr}(e^{-\beta H/2}e^{iHt}a_pe^{-iHt}a_q^{\dag}e^{-\beta H/2}) = \sum_n \bra{n}e^{-\beta H/2}e^{iHt}a_pe^{-iHt}a_q^{\dag}e^{-\beta H/2}\ket{n}.
\end{equation}
Indexing the eigenbasis of $H$ with the vectors $\ket{n}$, we obtain
\begin{equation}
    \mathrm{Tr}(e^{-\beta H/2}e^{iHt}a_pe^{-iHt}a_q^{\dag}e^{-\beta H/2}) = \sum_n \bra{n}e^{-\beta H/2}a_pe^{-i(H-E_n)t}a_q^{\dag}e^{-\beta H/2}\ket{n}.
\end{equation}
We now require the full spectrum of the Hamiltonian as input, but regardless, taking the Fourier transform we obtain 
\begin{equation}
    G_{pq}^+(\beta, \omega) = \sum_n e^{-\beta E_n} \bra{n}a_p\left[\omega - (H - E_n) \right]^{-1}a_q^{\dag}\ket{n}.
\end{equation}
In principle, this could be solved using the Arnoldi method if the full spectrum of the Hamiltonian was known by performing the Arnoldi iteration on every eigenstate $\ket{n}$. 
However, this is obviously infeasible for anything other than tiny systems.

We develop a different method utilizing the thermofield double (TFD) state~\cite{Harsha_2019}, defined as
\begin{equation}
    \ket{\Psi(\beta)} = \sqrt{\frac{1}{Z(\beta)}} \sum_i \left( \mathbb{1} \otimes e^{-\beta H/2}  \right) \ket{i}\ket{i},
    \label{eq:TFD}
\end{equation}
We do this by way of showing that the TFD state is a vectorization of the matrix $e^{-\beta H/2}$, and then transform the trace in Eq.~\ref{eq:therm_G(t)} with a series of vectorization identities.

\subsubsection{Vectorization} 

We first define the vectorization of a square complex matrix $\left(\mathrm{vec}: \mathbb{C}^{n\times n} \rightarrow \mathbb{C}^{2n} \right)$ as the map 
\begin{equation}
    \sum_{i,j} M_{ij} \ket{i}\bra{j} \rightarrow \sum_{i,j} M_{ij} \ket{j}\ket{i}.
    \label{eq:vec}
\end{equation}
The following identities are easily verified through direct calculation:
\begin{subequations}
\begin{align}
    \mathrm{vec}(AB) &= \left( \mathbb{1} \otimes A  \right) \mathrm{vec}(B),  \label{eq:vec1} \\
    \mathrm{vec}(AB) &= \left( B^T \otimes \mathbb{1}  \right) \mathrm{vec}(A), \label{eq:vec2}\\
    \mathrm{Tr}(A^{\dag}B) &= \mathrm{vec}(A)^{\dag}\mathrm{vec}(B). \label{eq:vec3}
\end{align}
\end{subequations}

Eqs.~\ref{eq:vec1} and~\ref{eq:vec2} imply we have freedom to place operators on either side of the tensor product and Eq.~\ref{eq:vec3} shows us we can evaluate the trace of operators as the inner product of their respective vectorizations.
Furthermore, from Eq.~\ref{eq:vec1} we recognize that the TFD state in Eq.~\ref{eq:TFD} is a vectorization of $e^{-\beta H/2}\mathbb{1} = e^{-\beta H/2}$
\begin{equation}
    \ket{\Psi(\beta)} = \mathrm{vec}(e^{-\beta H/2}).
    \label{eq:TFD_vec}
\end{equation}
We will use these to express the thermal Green's function using a single resolvent.

\subsubsection{Green's Function with TFD}

Using Eq.~\ref{eq:vec3}, the numerator in Eq.~\ref{eq:therm_G(t)} can be written as
\begin{equation}
    \mathrm{Tr}(e^{-\beta H/2}e^{iHt}a_pe^{-iHt}a_q^{\dag}e^{-\beta H/2}) = \mathrm{vec}(e^{-iHt}e^{-\beta H/2} )^{\dag}\mathrm{vec}(a_pe^{-iHt}a_q^{\dag}e^{-\beta H/2} ).
\end{equation}
Since $e^{-iHt}$ and $e^{-\beta H/2}$ commute, we can rearrange their ordering and use Eqs.~\ref{eq:vec1} and~\ref{eq:vec2} to place the leftmost time evolution operator on the other side of the tensor product as the other operators
\begin{equation}
    \mathrm{vec}(e^{-\beta H/2}e^{-iHt} )^{\dag}\mathrm{vec}(a_pe^{-iHt}a_q^{\dag}e^{-\beta H/2} ) = \mathrm{vec}(e^{-\beta H/2} )^{\dag}\left(e^{iHt} \otimes \mathbb{1} \right)\left(\mathbb{1} \otimes  a_pe^{-iHt}a_q^{\dag}\right)\mathrm{vec}(e^{-\beta H/2} ).
\end{equation}
Which we simplify using Eq.~\ref{eq:TFD_vec}
\begin{equation}
    \mathrm{vec}(e^{-\beta H/2} )^{\dag}\left(e^{iHt} \otimes \mathbb{1} \right)\left(\mathbb{1} \otimes  a_pe^{-iHt}a_q^{\dag}\right)\mathrm{vec}(e^{-\beta H/2} ) = \bra{\Psi(\beta)}\left(e^{iHt} \otimes a_pe^{-iHt}a_q^{\dag}\right)\ket{\Psi(\beta)}.
\end{equation}
Because they act on different qubits, the inverse time evolution operator now commutes with the annihilation operator and we write
\begin{equation}
    \bra{\Psi(\beta)}\left(e^{iHt} \otimes a_pe^{-iHt}a_q^{\dag}\right)\ket{\Psi(\beta)} = \bra{\Psi(\beta)} \left( \mathbb{1} \otimes a_p \right) \left(e^{i( H\otimes \mathbb{1} - \mathbb{1} \otimes H)t }\right) \left( \mathbb{1} \otimes a_q^{\dag} \right) \ket{\Psi(\beta)}.
\end{equation}
Now we obtain the Green's function in the frequency domain via a Fourier transform
\begin{align}
    G_{pq}(\beta, \omega)^+ &= \int_{-\infty}^{\infty}dt e^{i\omega t}\Theta(t)\bra{\Psi(\beta)} \left( \mathbb{1} \otimes a_p \right) \left(e^{i( H\otimes \mathbb{1} - \mathbb{1} \otimes H)t }\right) \left( \mathbb{1} \otimes a_q^{\dag} \right) \ket{\Psi(\beta)} \nonumber \\
    &= \bra{\Psi(\beta)} \left( \mathbb{1} \otimes a_p \right) \int_{0}^{\infty}dt\left(e^{i\omega t)}e^{i( H\otimes \mathbb{1} - \mathbb{1} \otimes H)t }\right) \left( \mathbb{1} \otimes a_q^{\dag} \right) \ket{\Psi(\beta)} \nonumber \\
    &= \bra{\Psi(\beta)} \left( \mathbb{1} \otimes a_p \right) \left[ \omega + ( H\otimes \mathbb{1} - \mathbb{1} \otimes H)\right]^{-1} \left( \mathbb{1} \otimes a_q^{\dag} \right) \ket{\Psi(\beta)}.
\end{align}

The same analysis applies to the backwards Green's function, thus we state
\begin{subequations}
\begin{align}
    G_{pq}^+(\beta, \omega) &= \bra{\Psi(\beta)} \left( \mathbb{1} \otimes a_p \right) \left[ \omega + ( H\otimes \mathbb{1} - \mathbb{1} \otimes H)\right]^{-1} \left( \mathbb{1} \otimes a_q^{\dag} \right) \ket{\Psi(\beta)}  \\
    G_{pq}^-(\beta, \omega) &= \bra{\Psi(\beta)} \left( \mathbb{1} \otimes a_p^{\dag} \right) \left[ \omega - ( H\otimes \mathbb{1} - \mathbb{1} \otimes H)\right]^{-1} \left( \mathbb{1} \otimes a_q \right) \ket{\Psi(\beta)}.
    \label{eq:G_therm_sm}
\end{align}
\end{subequations}
Thus we have a means to use the Arnoldi method to evaluate the Green's function at $T > 0$.
Namely we use the same protocol as the zero-temperature case but substitute $\ket{\Psi(\beta)}$ of $\ket{\psi_0}$ and $( H\otimes \mathbb{1} - \mathbb{1} \otimes H)$ for $H$.

\subsection{The spectral function}

For small $\gamma$, the imaginary component of $G(\omega)$ describes the probability distribution of single-particle excitations.
This is called the spectral function
\begin{equation}
    A_{pq}(\omega,\gamma) = -\pi^{-1}\text{Im}(G_{pq}(\omega+i\gamma)),
\end{equation}
It will be useful to consider the effects of $\gamma$ when estimating the spectral function, therefore we explicitly account for it in our definition.

\section{Estimating the Green's function with QSVT}
\label{sec:QSVT}

Here we explain how Green's functions could be estimated using the quantum singular value transformation.
We assume a basic understanding of block-encoding ~\cite{Low_2019, Chakraborty_2019} and QSVT~\cite{Gily_n_2019} and refer the reader to the respective citations as well as Ref.~\cite{lin2022lecturenotesquantumalgorithms} chapters six and eight for more details on these methods.
Starting from a linear combination of unitaries (LCU)~\cite{childs2012hamiltonian} block encoding of $H$, we form a block encoding of $(\omega + i\gamma \mp E_0)\mathbb{1} \pm H $ simply by adding an identity term to the LCU decomposition of $H$.
We then use QSVT to transform this block encoding into a polynomial approximating $1/x$, yielding a block encoding of $\left[ (\omega \mp E_0)\mathbb{1} \pm H \right]^{-1}$.

\subsection{The matrix inversion polynomial}

We now show the polynomial used for matrix inversion following Refs.~\cite{childs2017quantum} and~\cite{Martyn_2021}.
Like all methods for matrix inversion the complexity depend on the matrix's condition number $\kappa$.
Here $\kappa$ is an \textit{effective} condition number for the block-encoded matrix given as the ratio of the subnormalization factor to the eigenvalue of smallest absolute value 
\begin{equation}
    \kappa = \frac{\lambda}{\sigma_{\textrm{min}}},
    \label{eq:eff_cond}
\end{equation}
which is lower bounded by the actual condition number of the block-encoded matrix.

Setting $b = \lceil \kappa^2\log{(\kappa/\epsilon)} \rceil$, the function
\begin{equation}
    g^{\textrm{inv}}(x) = \frac{1-\left(1 -x^2\right)^b}{x}.
\end{equation}
provides an $\epsilon$-approximation to $1/x$ on the range $1/\kappa \leq |x| \leq 1$.
This function $g^{\textrm{inv}}(x)$ is approximated by the polynomial
\begin{equation}
    p^{\textrm{inv}}(x) = 4\sum_{j=0}^d (-1)^j \left[ 2^{-2b}\sum_{i=j+1}^b {2b \choose b+i} \right]T_{2j+1}(x)
\end{equation}
by setting $d = \left\lceil \sqrt{b\log(4b/\epsilon)}  \right\rceil$, where $T_{2j+1}(\cdot)$ denotes the $(2j+1)$th Chebyshev polynomial.
$p^{\textrm{inv}}(x)$ is then an $2\epsilon$ approximation to $1/x$ on the range $1/\kappa \leq |x| \leq 1$.
Unfortunately, this polynomial cannot yet be used as it is not bounded by $1$ between $-1/\kappa < x < 1/\kappa$ and therefore cannot be realized as a QSVT polynomial. 
In order to tame the behavior of this polynomial near the origin, we will multiply by a rectangle function equal to $1$ for $1/\kappa \leq |x| \leq 1$ and $0$ for $|x| \leq 1/\kappa$.
As the rectangle function is the sum of two step functions it can be approximated by a polynomial approximating the sum of two error functions~\cite{sachdeva2013approximationtheorydesignfast}.
This polynomial is given as 
\begin{equation}
    p^{\textrm{rect}}(x) = \frac{2ke^{-k^2/2}}{\sqrt{\pi}} \left( I_0(k^2/2)(x-\delta) + \sum_{j=1}^{(n-1)/2}(-1)^jI_j(k^2/2)\left(\frac{T_{2j+1}(x-\delta)}{2j+1} - \frac{T_{2j-1}(x-\delta) }{2j-1}\right) \right).
\end{equation}
Where $I_j(\cdot)$ denotes the $j$th Bessel function of the first kind.
The approximation error on this function is
\begin{equation}
    \epsilon_{\textrm{rect}} \leq \frac{4k}{\sqrt{\pi}n}\left(e^{-n^2/8t} + e^{-k^2/2-t}\right).
\end{equation}
To achieve error $\epsilon_{\textrm{rect}}$, we set $t = \lceil \max(\beta e^2, \log(4/\epsilon_{\textrm{rect}}) ) \rceil $ and $n = \lceil\sqrt{2t\log{(4/\epsilon_{rect})} }  \rceil$.

The overall QSVT matrix inversion polynomial is the product of the inversion and rectangle polynomials
\begin{equation}
    p^{\textrm{MI}}(x) = \frac{1}{2\kappa}p^{\textrm{inv}}(x)p^{\textrm{rect}}(x).
    \label{eq:mi_poly}
\end{equation}
If we approximate $p^{\textrm{inv}}$ to $\epsilon_{\textrm{qsvt}}$ and $p^{\textrm{rect}}$ to $\min[2\epsilon_{\textrm{qsvt}}/5\kappa, \kappa/2d ]$, then
\begin{equation}
    \left|p^{\textrm{MI}}(x) - \frac{1}{2\kappa}\frac{1}{x}\right| \leq \frac{\epsilon_{\textrm{qsvt}}}{2\kappa}.
\end{equation}
The degree of $p^{\textrm{MI}}(x)$ is $d + n$, where $d$ and $n$ are set as
\begin{subequations}
\begin{align}
    b &= \lceil \kappa^2 \log(\kappa/\epsilon_{\textrm{qsvt}}) \rceil \\
    d &= \lceil \sqrt{b\log(4b/\epsilon_{\textrm{qsvt}}) } \rceil \\
    \epsilon_{\textrm{rect}} &= \min[2\epsilon_\textrm{{qsvt}}/5\kappa, \kappa/2d ] \\
    t &= \lceil \max(\beta e^2, \log(4/\epsilon_\textrm{{rect}}) ) \rceil \\
    n &= \lceil\sqrt{2t\log{(4/\epsilon_\textrm{{rect}})} }  \rceil.
\end{align}
\label{eq:degree}
\end{subequations}

The polynomial formed by plugging these parameters into Eq.~\ref{eq:mi_poly} now transforms our block-encoding of $(\omega \mp E_0)\mathbb{1} \pm H $ with subnormalization factor $\lambda_H + |\omega \mp E_0|$ into a block-encoding of $\left[ (\omega \mp E_0)\mathbb{1} \pm H \right]^{-1}$ with subnormalization factor $2/\sigma_{\textrm{min}}$ and error $\epsilon_{\textrm{qsvt}}/(2\kappa)$.
That is
\begin{equation}
    \left|p^{\textrm{MI}} \left( \frac{ (\omega +i\gamma \mp E_0)\mathbb{1} \pm H}{\lambda_H + |\omega + i\gamma \mp E_0|} \right) - \left( \frac{\sigma_\textrm{{min}}}{2} \right) \frac{1}{(\omega + i\gamma \mp E_0)\mathbb{1} \pm H}\right| \leq \frac{\epsilon_{\textrm{qsvt}}}{2\kappa}.
\end{equation}

\subsection{The Hadamard test for block-encoded matrices}

Using the matrix inversion polynomial we use QSVT to transform a block-encoding of $\omega +i\gamma \pm (H - E_0)$ into a block-encoding of $\left[ \omega + i\gamma \pm (H - E_0) \right]^{-1}$, what remains is to estimate the expectation value of this matrix with respect to the $a_p\ket{\psi_0}$ or $a_p^{\dagger}\ket{\psi_0}$ states.
The Hadamard test is a well-known protocol to estimate the expectation values of unitaries matrices.
It is easily extended to estimate the expectation value of a block-encoded matrix~\cite{Tong_2021}.
That is, for a matrix $A$ block-encoded into the unitary $U_A$ using $m$ ancilla qubits, the expectation value 
\begin{equation}
    \bra{0}^{\otimes m}\bra{\Psi} U_A \ket{0}^{\otimes m}\ket{\Psi} = \braket{\Psi|A|\Psi}.
\end{equation}
Therefore, we simply prepare the ancilla register to $\ket{0}^{\otimes m}$ and preform a Hadamard test of $U_A$ on the system plus ancilla registers.

\subsection{An \textit{a posteriori} bound for the ROQAM}

The motivation for the ROQAM for Green's function estimation is to avoid needing to employ a seperate instance of QSVT or QLSA for every frequency $\omega$.
However, one of the main drawbacks of the ROQAM is the lack of tight $\textit{a priori}$ bounds under finite-precision estimation.
An \textit{a posteriori} bound can be established if we are allowed to use QSVT to estimate $G(\omega)$ at a single $\omega$ that ideally well represents the overall difficulty of estimating these functions (e.g., near a pole on the real axis or at the truncation frequency on the imaginary axis).
The QSVT estimate will be taken to some precision $\Delta_{QSVT}$.
Then take the difference between estimate from the ROQAM at that same point is compared to the estimate obtained via QSVT to be $\Delta_{diff}$.
Then the error on the estimate obtained via the ROQAM will be bounded as
\begin{equation}
    \Delta_{ROQAM} \leq \Delta_{QSVT} + \Delta_{diff}.
\end{equation}
This \textit{a posteriori} bound could be used during iteration to guide selection of $r$.
Furthermore, if restarts are employed, this could be used to tune $\Delta t$ and the $\delta_l$'s at the cost of some additional overhead.

\section{Numerical Tests}
\label{sec:numbers}

Here, we describe numerical tests of the ROQAM applied to the estimation of single-particle Green's functions for instances of the single-impurity Anderson model (SIAM).

\subsection{The single-impurity Anderson model}

The SIAM is a simple model of a strongly correlated system, describing a localized quantum impurity coupled to a bath of free fermions.
The SIAM Hamiltonian given $n_{bath}$ spin orbitals is defined as
\begin{align}
    H &= H_{imp} + H_{bath} + H_{int} \nonumber \\
    H_{imp} &= \sum_{\sigma}(\varepsilon_{imp} - \mu)a^{\dagger}_{\sigma}a_{\sigma} + Ua^{\dagger}_{\downarrow}a_{\downarrow}a^{\dagger}_{\uparrow}a_{\uparrow} \nonumber \\
    H_{bath} &= \sum_{\sigma, j = 1}^{n_{bath}} (\varepsilon_j-\mu)c^{\dagger}_{j\sigma}c_{j\sigma}  \nonumber \\
    H_{int} &= \sum_{\sigma, j = 1}^{n_{bath}} V_j \left( a^{\dagger}_{\sigma}c_{j\sigma} + c^{\dagger}_{j\sigma}a_{\sigma} \right).
\end{align}
The action of the bath is to hybridize the impurity orbitals with the bath orbitals representing a conduction band, adding a finite width to the impurity levels.
This action is fully captured through the hybridization function~\cite{RevModPhys.68.13}
\begin{equation}
    \Delta(\omega) = \sum_j^{n_{bath}} \frac{V_j^2}{\omega + i\gamma - (\varepsilon_j - \mu)}.
    \label{eq:hyb}
\end{equation}

In the limit $n_{bath} \rightarrow \infty$, the Hubbard model exactly maps onto the SIAM, making the SIAM an entry point for dynamical mean field theory (DMFT)~\cite{RevModPhys.68.13}, a powerful framework for the study of strongly correlated materials.
In order for this limit to meaningful the hopping terms must be scaled like
\begin{equation}
    V_j = \mathcal{O}\left(\frac{1}{\sqrt{n_{bath}}} \right),
    \label{eq:scale}
\end{equation}
such that the hybridization function described by Eq.~\ref{eq:hyb} converges.
In DMFT, a lattice problem is mapped onto an impurity problem, either the SIAM or one with multiple impurity orbitals.
The bath sites are then parameterized through a self-consistency loop such as to best represent the local physics of the originating lattice problem.

The smallest instance of DMFT is known as ``two-site DMFT''~\cite{Potthoff_2001}, where $n_{bath}$ is fixed to $1$.
Formulated using the single-band Hubbard model at half-filling ($\mu = U/2$) on the Bethe lattice with infinite connectivity, the self-consistency conditions can be analytically solved yielding
\begin{subequations}
    \begin{align}
        \varepsilon_1 &= \mu \\
        V_{1} &= \sqrt{1 - \frac{U^2}{36}}
    \end{align}
\end{subequations}
For $U \geq 6$ the hopping term is $V_{1}=0$.
This gives a rough picture of the Mott transition where at $U < 6$ the SIAM behaves like a metal while for $U \geq 6$ the SIAM is an insulator.

The SIAM we study uses this two-site solution as a starting point.
We reuse these parameters for all bath sites with the hopping terms rescaled as Eq.~\ref{eq:scale}
\begin{subequations}
    \begin{align}
        \varepsilon_j &= \varepsilon_1 \\
        V_{j} &= \frac{V_{1}}{\sqrt{n_{bath}}}
    \end{align}
\end{subequations}
This could be then seen as a crude initial guess for a larger DMFT instance where the SIAM contains multiple bath sites.

We focus on the diagonal elements of the impurity site Green's function.
At zero temperature this is defined using Equation~\ref{eq:G(w)_sm} as
\begin{align}
     G^+_{\sigma}(\omega) &= \bra{\psi_0}a_{\sigma}\left[ \omega - (H - E_0) \right]^{-1}a_{\sigma}^{\dag}\ket{\psi_0} \nonumber \\
     G^-_{\sigma}(\omega) &= \bra{\psi_0}a_{\sigma}^{\dag}\left[ \omega + (H - E_0) \right]^{-1}a_{\sigma}\ket{\psi_0} \nonumber \\
     &G_{\sigma}(\omega) = G^+_{\sigma}(\omega) + G^-_{\sigma}(\omega)
\end{align}

In our tests, the bath size was limited to $n_{bath} \leq 5$, allowing for verification of estimates through comparison to direct matrix inversion.
These small system sizes allowed us to obtain the ground state eigenpair $\left( \ket{\psi_0}, E_0 \right)$ exactly, so as to study the estimation error resulting from the Arnoldi method alone and not from noisy inputs.

\subsection{Emulating the ROQAM}

\begin{figure}[!h]
    \includegraphics{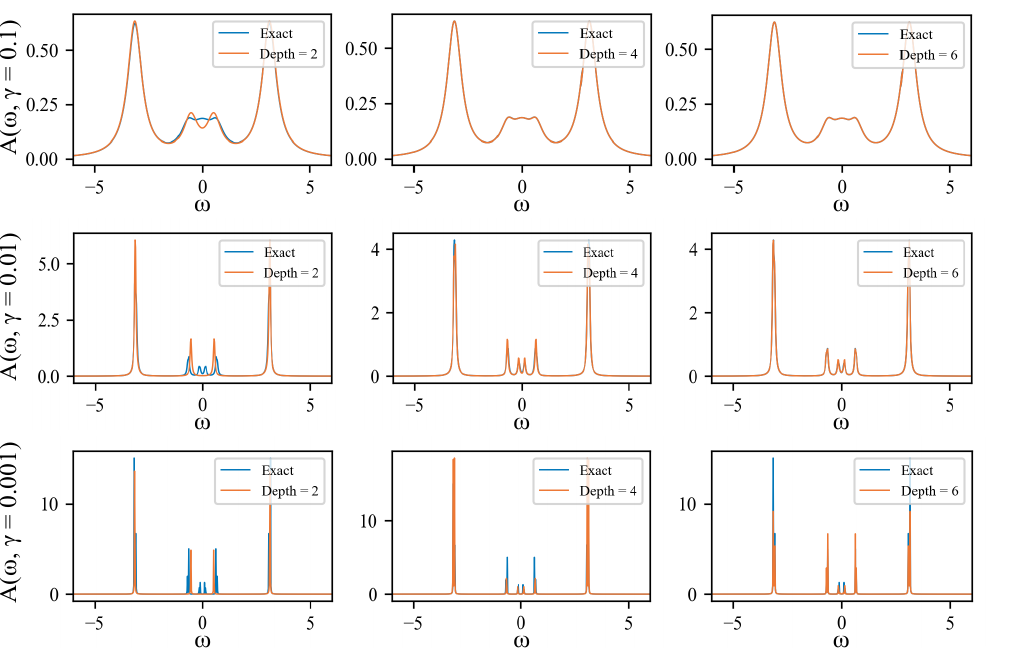}
    \caption{Estimates of the spectral function of SIAM with $n_{bath} = 4$ with $\gamma = 0.1$ (top), $0.01$ (middle) and $0.001$ (bottom) in units of the Hubbard bandwidth. Each iterate adds $2$ poles to the estimate of $A(\omega,\gamma)$. When the poles of $A(\omega, \gamma)$ are clustered near each other, increasing $\gamma$ will combine these poles, requiring fewer iterates of the ROQAM to well approximate}
    \label{fig:spectral_converge}
\end{figure}

To estimate $G^+_\sigma(\omega)$ at zero temperature we performed the Arnoldi iteration on the starting state $\ket{\chi_0} = a_{\sigma}\ket{\psi_0}$ using the generating matrix $U = e^{-iH\Delta t}$.
We used a Python program to calculate the expectation values $\bra{\chi_0}U^l\ket{\chi_0}$.
To emulate the effects of finite sampling, independently random numbers with mean zero and variance $\delta^2_l$ were added to the real and imaginary components of these expectation values.
We then constructed $[U]$ from Equation~\ref{eq:poly_matrix_elements}.
The matrix $[U]$ is then processed into $[H]$ using Equation~\ref{eq:H_from_U}.
Estimates of $G^+_\sigma(\omega)$ were then formed as
\begin{align}
    \hat{G}_{r, \sigma}^+(\omega) = \left[ \left( (\omega + E_0)\mathbb{1}_r - [H] \right)^{-1} \right]_{00}.
\end{align}
The same method was used to estimate $G^-_\sigma(\omega)$, simply changing the starting state to $\ket{\chi_0} = a_{\sigma}^{\dagger}\ket{\psi_0}$.
Once estimates of both $G^+_\sigma(\omega)$ and $G^-_\sigma(\omega)$ were obtained, we simply added them together to estimate $G_{\sigma}(\omega)$.
We always performed the iteration to the same depth for both $G^+_\sigma(\omega)$ and $G^-_\sigma(\omega)$ (unless breakdown occurred for one but not the other), thus our estimate for $G_{\sigma}(\omega)$ given an iteration depth $r$ is
\begin{equation}
    \hat{G}_{r, \sigma}(\omega) = \hat{G}_{r, \sigma}^+(\omega) + \hat{G}_{r, \sigma}^-(\omega)
\end{equation}
Once $\hat{G}_{r, \sigma}(\omega)$, we estimate the spectral function as
\begin{equation}
    \hat{A}_{r, \sigma}(\omega, \gamma) = -\frac{1}{\pi}\hat{G}_{r, \sigma}(\omega + i\gamma) 
\end{equation}
Before considering the impact of finite-sampling errors, we first studied the convergence of the method in the ideal noiseless case.
Here ``noiseless'' means we did not add random noise to mimic the effects of finite sampling errors, all computations were performed using with numpy’s double-precision floating-point arithmetic.
In Fig.~\ref{fig:spectral_converge}, we show the convergence of the spectral function for the SIAM studied with $4$ bath spin-orbitals and onsite interaction $U = 5$ with three different values of the broadening parameter $\gamma$.
Although it enters after the iteration, $\gamma$ greatly impacts the overall convergence rate with iteration depth.
At low $\gamma$, the ROQAM is simply being used to find all the eigenvalues of $H$ with support on $\ket{\chi_0}$.

\begin{figure}[!h]
    \centering
    \includegraphics{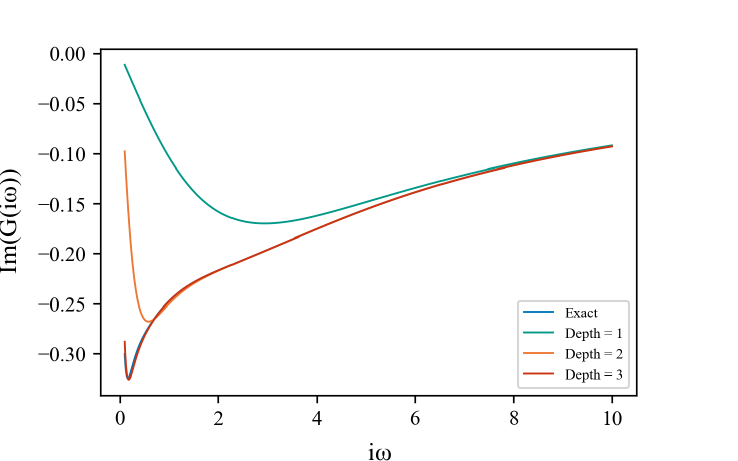}
    \caption{The estimates $\text{Im}(\hat{G}(i\omega))$ on the imaginary frequency axis (denoted $i\omega$). Because the poles of the Green's function lie solely on the real frequency axis $G(i\omega)$ is always smooth and continuous without the need for a broadening parameter. We see that after only $3$ iterations $\text{Im}(\hat{G}(i\omega))$ begin to well approximate $\text{Im}(G(i\omega))$.}
    \label{fig:convergence_on_im_axis}
\end{figure}

In some cases, such as DMFT, the Green's function on the imaginary frequency axis is relevant.
This is denoted as $G(i\omega)$.
On the imaginary axis, the Green's function is continuous and smooth everywhere with no poles.
The convergence is then similar to that of the spectral function broadened to form a continuous spectra.
This has the advantage of obtaining the faster convergence without the trade-off of decreasing spectral resolution.
In Figure~\ref{fig:convergence_on_im_axis} we show the convergence of $\hat{G}_{r}(i\omega)$ on the imaginary axis.
We see that after only $3$ iterations $\hat{G}_{r}(i\omega)$ begins to well approximate $G(i\omega)$.

\begin{figure}[!h]
    \centering
    \includegraphics{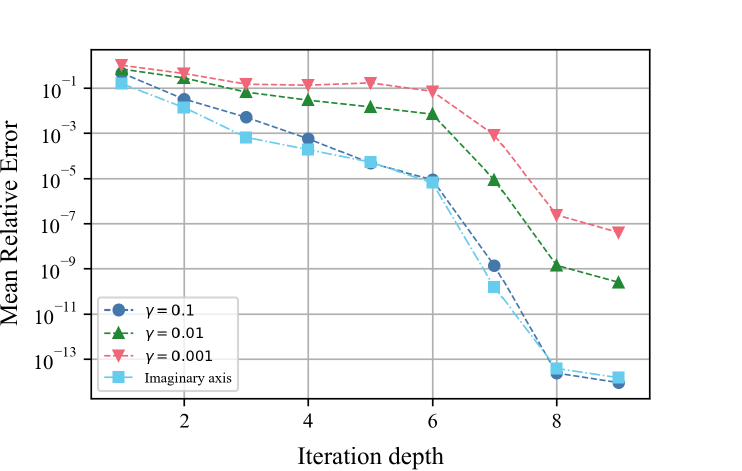}
    \caption{The mean relative error on the spectral function vs ROQAM iteration depth with three values of $\gamma$. In limit $\gamma \rightarrow 0$, the error does not improve until all the peaks of $A(\omega)$ have been located. This is equivalent to finding all eigenvalues of $H$ with support on $\ket{\chi_0}$. In contrast, when $\gamma$ is set such that $A(\omega)$ is continuous every iterate steadily improves the error.
    We also show the mean relative error of $G(i\omega)$ on the imaginary axis.
    Convergence on the imaginary axis is very similar to spectral function with $\gamma = 0.1$.}
    \label{fig:errors_v_depth}
\end{figure}

In Fig.~\ref{fig:errors_v_depth}, we show the mean relative error of $\hat{G}_r(\omega)$ with $\gamma = 0.1$, $0.01$, and $0.001$ in units of the Hubbard bandwidth. 
For low $\gamma$, the ROQAM must find the locations of all the poles of $G(\omega)$ before further iterates improve the errors.
This is seen where the mean relative error flatlines for $\gamma = 0.001$ at intermediate iterates before improving after the sixth iteration.
In contrast, when $\gamma = 0.1$, $A(\omega, \gamma)$ is a continuous function and the ROQAM determines its approximate shape very quickly steadily improving the error with each iterate.
We also show the mean relative error of $\hat{G}_r(i\omega)$ on the imaginary axis.
The imaginary axis behaves similar to the real axis when $\gamma = 0.1$, where every iterate improves the error.

\begin{figure}[!h]
    \centering
    \includegraphics{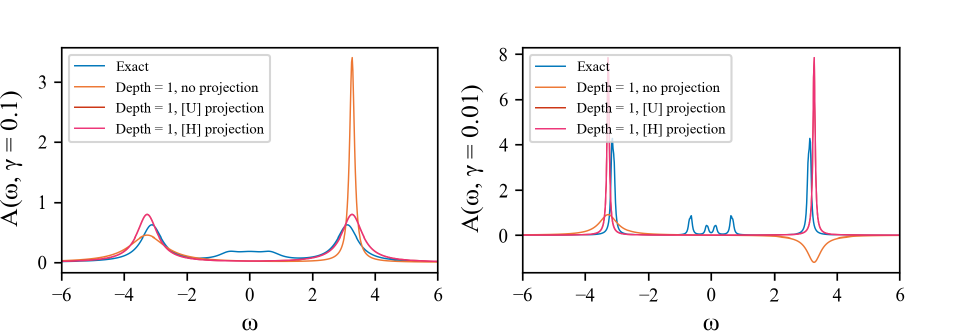}
    \caption{Example of an artifact resulting from low iteration depth. The Arnoldi iteration is performed to depth $1$ on a SIAM at half-filling with $4$ bath spin-orbitals. For both $\gamma = 0.1$ and $0.01$, the estimate $\hat{G}_1(\omega)$ locates the approximate positions of two poles, but contains obviously errors. At $\gamma = 0.1$, the spectral function appears distorted and asymmetrical while at $\gamma = 0.1$ the pole carries a negative sign. This can be resolved by either by projecting $[U]$ to the nearest unitary or projecting the $[H]$ obtained from Equation~\ref{eq:H_from_U} to the nearest Hermitian matrix. The estimates resulting from these two choices are nearly indistinguishable.}
    \label{fig:low_depth_sign_error}
\end{figure}

At low depths, the estimates from the Arnoldi method $\hat{G}_r(\omega)$ can display erratic behavior near the real frequency axis.
This is an artifact which arises because while the matrix $[U]$ represents the unitary $U$ in $\mathcal{K}_r(U,\ket{\chi_0})$, $[U]$ is not generally unitary itself for $r < r_0$.
This implies that the matrix $[H]$ obtained from Equation~\ref{eq:H_from_U} is not quite Hermitian, containing eigenvalues with small nonzero imaginary components.
As a result, the poles of $\hat{G}_r(\omega)$ are not exclusively on the real axis, as is the case for the actual Green's function.
An example is shown in Figure~\ref{fig:low_depth_sign_error}, where the Arnoldi iteration is only performed to depth $1$.
The resulting $\hat{G}_r(\omega)$ at $\gamma = 0.1$ appears distorted  and at $\gamma = 0.01$ has the wrong sign near one of the poles of $G(\omega)$.
This can be handled in a few different ways.
First, it could be simply ignored as a low depth effect, $\hat{G}_r(\omega)$ will still provide qualitatively correction information about the locations of peaks.
As $r$ increases $[U]$ becomes closer to a unitary and this issue resolves itself.
Alternatively, this effect can be removed manually by enforcing that matrix $[H]$ be Hermitian.
This can be done either by projecting $[U]$ to the closest unitary, or by projecting $[H]$ to the closest Hermitian.
The closest unitary in $l_2$ norm distance is obtained by taking the singular value of decomposition of $[U]$ and discarding the diagonal matrix
\begin{equation}
    [U] = P\Sigma V^{\dagger} \rightarrow PV^{\dagger}.
\end{equation}
Likewise, the closest Hermitian matrix to $[H]$ is obtained by averaging $[H]$ with its Hermitian conjugate
\begin{equation}
    [H] \rightarrow \frac{1}{2}\left( [H] + [H]^\dagger  \right).
\end{equation}
In the absence of noise, these two projections produce nearly indistinguishable $[H]$ matrices.

\subsection{Behavior under finite-precision errors}
\label{sec:finite-precision_errors}

We also studied the behavior of the ROQAM in the presence of finite-precision estimation.
To mimic estimating the expectation values $\braket{\chi_0|U^l|\chi_0}$ to precision $\delta_l$, we add independent random variables with standard deviation $\delta_l$ to the real and imaginary components of $\braket{\chi_0|U^l|\chi_0}$.
The main effect of finite precision will be to bias the estimator $\hat{G}_r(\omega)$.
However, finite-precision errors have unique interactions with various aspects of the quantum Arnoldi iteration, which we will now elaborate on.

\begin{figure}[!h]
    \centering
    \includegraphics{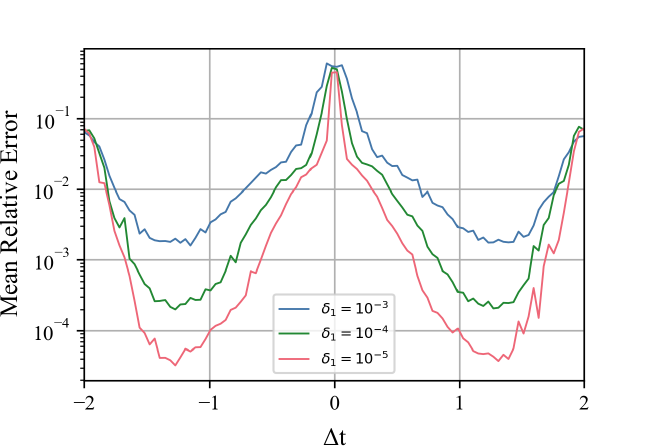}
    \caption{Dependence of the error on the choice of timestep  for a SIAM with $4$ bath sites on the real and imaginary axes for two different choices of estimation precision. Near zero, finite precision errors are greatly amplified across iterates. There is a sharp edge at the no-aliasing condition, where when $\Delta t$ is too large. The optimal choice of $\Delta t$ is that which minimizes the forward amplification of finite precision errors.}
    \label{fig:timesteps}
\end{figure}

As discussed in Section~\ref{sec:delta_t}, the selection of $\Delta t$ becomes nuanced in finite precision.
In Fig.~\ref{fig:timesteps} we show results for varying $\Delta t$ when performing the Arnoldi iteration to finite precision.
There is a strict maximum for $|\Delta t|$ set by the no-aliasing condition (Eq.~\ref{eq:no_aliasing_sm}), we see a sharp transition in error when $|\Delta t|$ is above this critical value.
Additionally, errors at every step of the iteration get amplified by a factor of $||\ket{\tilde{\chi}_j}||^{-1}$ when $\ket{\tilde{\chi}_j}$ is normalized.
When $\Delta t$ is near zero, the time evolution operator is very close to the identity matrix meaning the norms of the Arnoldi states $\ket{\tilde{\chi}_j}$ will be small.
Thus for small $\Delta t$ there is severe error amplification across iterates, rapidly degrading the estimate $\hat{G}_r(\omega)$.
The error improves when $\Delta t$ is moved away from zero, however, the exact dependence is very complicated and is not always monotonically improving with increasing $|\Delta t|$.
The optimal value of $|\Delta t|$ can be below the no-aliasing condition.

In the noiseless case we pointed out that $[H]$ is not quite Hermitian before the Krylov subspace stabilizes.
This effect is magnified in the presence of finite-sampling errors.
We find that performing the projection improves the estimator error by a factor of $\sim (1-2)$.
Again it is negligible whether this is done by projecting $[U]$ onto the closest unitary or projecting $[H]$ onto the closest Hermitian matrix.

\begin{figure}[!h]
    \centering
    \includegraphics{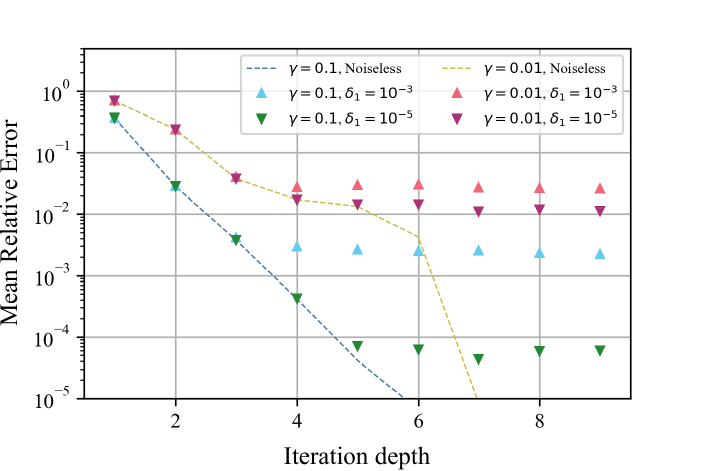}
    \caption{The error resulting from ROQAM with two levels of sampling precision (triangles) compared to the noiseless case (dashed line).
    When deployed in finite precision the estimator error from the ROQAM generally follows the noiseless error curve until hitting a noise floor set by the level of precision. }
    \label{fig:error_bias}
\end{figure}

The main effect of finite precision is a bias in the estimator $\hat{G}(\omega)$.
This is shown in Fig.~\ref{fig:error_bias}, we observe that the noisy estimates follow the noiseless curve until they hit this bias at which point further iterates do not improve $\hat{G}(\omega)$.

\begin{figure}[!h]
    \centering
    \includegraphics{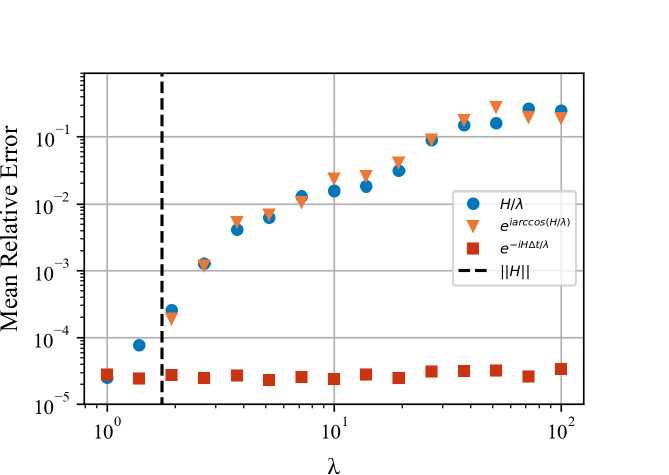}
    \caption{Error scaling vs block-encoding subnormalization factor $\lambda$ for a SIAM with $n_{bath} = 4$ and $\delta_1 = 10^{-5}$ with three different choices of generating matrix. Entries for the qubitized walk operator $(e^{i\arccos{(H/\lambda)} })$ begin after the dashed line because $\lambda$ is lower bounded by $||H||$. 
    When allowing a $\mathcal{O}(\lambda)$ increase in resources, the adverse effects of $\lambda$ are canceled out for the time evolution operator but not for the Hamiltonian or the qubitized walk operator.
    }
    \label{fig:lambda_scaling}
\end{figure}

In Fig.~\ref{fig:lambda_scaling} we show the error scaling with block-encoding subnormalization factor $\lambda$ for three choices of generating matrix: a block encoding of $H/\lambda$, the qubitized walk operator $e^{i\arccos(H/\lambda)}$, and the time evolution operator.
Each expectation value was estimated to $\delta = 10^{-5}$.
In each case we allow a linear increase in resources with $\lambda$.
For the time evolution operator this is spent rescaling $\Delta t' \rightarrow \lambda \Delta t$, this counteracts the adverse effects of $\lambda$ resulting.
In contrast, for $H/\lambda$ and $e^{i\arccos(H/\lambda)}$ we rescale the precision $\delta' \rightarrow \delta/\lambda$, this does not counteract $\lambda$, with a large increase in error even when $\lambda$ is set to its lower bound $||H||_2$.

In Section~\ref{sec:precision}, we pointed out that the all expectation values $\braket{\chi_0|U^l|\chi_0}$ do not need to estimated to the same precision. 
We studied three heuristics for setting the precisions.
The simplest choice is set $\delta$ to be on the order of the desired precision of $\hat{G}(\omega)$ and estimate each $\braket{\chi_0|U^l|\chi_0}$ to $\delta$.
\begin{equation}
    \delta_l = \delta
\end{equation}

We could also set the precisions based off how many times each expectation value is used.
The matrix $[U]$ is a $r \times r$ upper Hessenberg, meaning it contains
\begin{equation}
    \frac{1}{2}r(r+3)-1
\end{equation}
nonzero elements, with $\braket{\chi_0|U|\chi_0}$ appearing in all of them.
From inspection of Equations~\ref{eq:poly_matrix_elements} and~\ref{eq:auto_corrs}, we see that $\braket{\chi_0|U^l|\chi_0}$ does not appear in the $(j < l)$ columns of $[U]$.
This means that $\braket{\chi_0|U^l|\chi_0}$ appears in
\begin{equation}
    \frac{r^2+3r-l^2-l}{r^2+3r-2}
\end{equation}
elements of $[U]$.
This suggests a more economical error budget where we set $\delta_1$ to be on the order of the desired precision of $\hat{G}(\omega)$, estimate $\braket{\chi_0|U|\chi_0}$ to precision $\delta_1$, then estimate each subsequent $\braket{\chi_0|U^l|\chi_0}$ to precision
\begin{equation}
    \delta_l = \left(\frac{r^2+3r-2}{r^2+3r-l^2-l}\right)\delta_0.
    \label{eq:EB2}
\end{equation}

The final $\delta_r$ is then $\approx \delta_1/r$.
We also studied an aggressive error budget where each $\delta_l$ is scaled similarly.
\begin{equation}
    \delta_l = \frac{\delta_1}{l}.
\end{equation}

\begin{figure}[!h]
    \centering
    \includegraphics{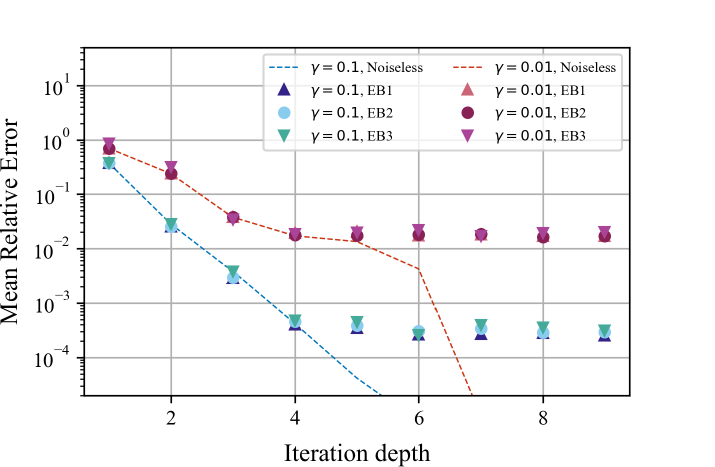}
    \caption{Comparison of three different error budgets. Error budget one (EB1) sets all $\delta_l$ to the same value $\delta$, error budget two (EB2) sets $\delta_l$ according to Eq.~\ref{eq:EB2}, where the precision is scaled to the number of times the estimate appears in $[U]$, finally error budget three (EB3) scales $\delta_l = l\delta_1$.
    We observe minimal difference in overall estimator precision between these three error budgets.}
    \label{fig:error_budgets}
\end{figure}

We compare these choices of error budgets in Figure~\ref{fig:error_budgets}.
We observe very little loss in estimator precision between these error budgets.

\subsection{Resource estimation}
\label{sec:QRE}

We estimate the $T$ gates required to estimate the SIAM $G(\omega)$ to $1 \%$ mean relative error on the real and imaginary axes at zero temperature.
These resource estimates do not include the cost of state preparation or the cost to estimate $E_0$, which is a input to the problem.
As such, these resource estimates should not be seen as precise estimates of the cost of the ROQAM, but rather a loose indicator of the relative difficulties of estimating $G(\omega)$.

We construct $U$ using a Trotter product. 
Because the state $\ket{\chi_0}$ is not an eigenstate of $U$, we cannot apply direction control to implement controlled-$U$ with only a single call to $U$.
Instead, controlled-$U$ will cost two queries to $U$. 
This effectively means doubling the number of Trotter steps.

We use iterative quantum amplitude estimation (IQAE)~\cite{Grinko_2021} to implement the Hadamard tests of $U$ with Heisenberg scaling.
This requires constructing a generalized Grover iterate from the output of the Hadamard test circuit, which will consist of two calls to controlled-$U$.
Using the Clopper-Pearson estimator, IQAE uses on average
\begin{equation}
    N \leq \frac{0.8}{\epsilon_{qae}}\log \left( \frac{2}{p_{fail} \log \left( \frac{\pi}{4\epsilon_{qae}} \right)} \right)
\end{equation}
calls to the generalized Grover iterate to achieve additive error $\epsilon_{qae}$ with failure probability $p_{fail}$.

We use the aggressive error budget across the quantum Arnoldi iterates explicitly calculate the iteration depth, $\Delta t$, and $\delta_0$ required to achieve $1 \%$ mean relative error on the real and imaginary axes.
The additive error $\delta_l$ is upper bounded by the sum of the IQAE estimation error, Trotter error, and rotation synthesis error
\begin{equation}
    \delta_l \leq \epsilon_{qae} + \epsilon_{trot} +\epsilon_{syn}.
\end{equation}

We optimize balancing the resources spent across these three sources of error by directly calculating the Trotter error
\begin{equation}
    \epsilon_{trot} = \braket{\chi_0| \left( U^l_{trot} - U^l \right)|\chi_0 }
\end{equation}
for given $\Delta t$ and number of Trotter steps.
We then calculate the total number of queries to $U_{trot}$ to achieve $\epsilon_{qae}$ using Clopper-Pearson IQAE.
From the number of Trotter steps and the number of queries to the Trotterized unitary we determine the total number of logical rotations to estimate $\braket{\chi_0|U^l|\chi_0}$ to precision $\epsilon_{qae} + \epsilon_{trot}$.
The remaining error is then spent synthesizing the logical rotations.
We utilize the the Kliuchnikov cost model~\cite{Kliuchnikov_2023} of 
\begin{equation}
    T_{rot} = 0.53\log(\epsilon_{syn}^{-1}) + 4.86
\end{equation}
$T$ gates per logical rotation. 
The synthesis error per rotation adds linearly in the worst case so we rescale the synthesis precision as
\begin{equation}
    \epsilon_{syn} \rightarrow \frac{\epsilon_{syn}}{N_{rots}}.
\end{equation}
For every step $l$, we scan over many choices of first-, second-, third-, or fourth-order Trotter product, number of Trotter steps, IQAE precision, and synthesis precision, we then select the combination with the lowest T cost.

We also estimate the resources to calculate the Green's function using a QSVT to perform matrix inversion on a block-encoding of $H$.
We construct the block-encoding using an LCU~\cite{LCU}.
Under a Jordan-Wigner transformation~\cite{jordan1928paulische} and neglecting identity terms, the SIAM Hamiltonian contains $L = 3 + 6n_{bath}$ Pauli terms.
We use logical rotations to form the $\textsc{prepare}$ oracle and unary iteration~\cite{Babbush_2018} to preform the $\textsc{select}$ oracle.
Thus $\textsc{prepare}$ costs $L - 1$ logical rotations and controlled $\textsc{select}$ costs $4L -4$ $T$ gates.
The block-encoding of $H$ requires two queries to $\textsc{prepare}$ and one to $\textsc{select}$ costing a total of $2L - 2$ rotations and $4L - 4$ $T$ gates.
The subnormalization factor is straightforwardly calculated as
\begin{equation}
    \lambda_H = \left(\epsilon_{imp} - \mu\right) + U +\sum_{j}^{n_{bath}}\left( \epsilon_j - \mu \right) + 2\sum_{j}^{n_{bath}}V_j.
\end{equation}
We add the term $(\omega \mp E_0)\mathbb{1}$ to this block-encoding slightly to instead block-encode $\omega \pm (H - E_0)$. 
This will add a single rotation to the $\textsc{prepare}$ oracle and no additional cost to $\textsc{select}$.
The subnormalization factor of this block-encoding is now
\begin{equation}
    \lambda = \lambda_H + |\omega \mp E_0|.
\end{equation}

We directly calculate the effective condition number (Eq.~\ref{eq:eff_cond}) at a given frequency. 
We then use Eqs.~\ref{eq:degree} to determine the required degree of the matrix-inversion polynomial.
Since the degree of the polynomial is equal to the number of queries to the block-encoding of $H$, it determines the resource cost of the QSVT.
The output of the QSVT is a block-encoding of $\left[ (\omega \mp E_0)\mathbb{1} \pm H \right]^{-1}$ with subnormalization factor $2/\sigma_{min}$.
Therefore, we must rescale the estimation precision by the new subnormalization factor.
Thus, to determine the $T$ counts required to estimate the Green's function within $1\%$ relative error using QSVT we calculate the requisite polynomial degree at every frequency $\omega$.
This is done while balancing the error resulting from  QSVT polynomial approximation error and QAE estimator error
\begin{equation}
    \epsilon \leq \frac{2}{\sigma_{min}}\epsilon_{qae} + \frac{1}{2\kappa}\epsilon_{qsvt}.
\end{equation}
We find prioritizing the IQAE estimator error over synthesis and polynomial approximation errors leads to the lowest $T$ counts.
We then estimate the total $T$ counts to estimate the Green's function on the real and imaginary frequency axes for $1000$ frequencies and for $100$ frequencies.
We also report the single frequency with the highest $T$ count to estimate.

\end{document}